\documentclass[journal]{IEEEtran}
\IEEEoverridecommandlockouts
\usepackage{cite}
\usepackage{amsmath,amssymb,amsfonts}
\usepackage{algorithmic}
\usepackage{graphicx}
\usepackage{textcomp}
\usepackage{xcolor}
\def\BibTeX{{\rm B\kern-.05em{\sc i\kern-.025em b}\kern-.08em
    T\kern-.1667em\lower.7ex\hbox{E}\kern-.125emX}}
    
\usepackage{subfigure}
\usepackage{bm}
\usepackage{amsmath}
\usepackage{multirow}
\usepackage[ruled]{algorithm2e}
\usepackage{float}
\usepackage{colortbl}
\usepackage{hyperref}
\usepackage[switch]{lineno}
\newtheorem{problem}{Problem}
\newtheorem{definition}{Definition}

\newcommand{\squishlist}{
   \begin{list}{$\bullet$}
    { \setlength{\itemsep}{0pt}      \setlength{\parsep}{3pt}
      \setlength{\topsep}{3pt}       \setlength{\partopsep}{3pt}
      \setlength{\leftmargin}{1em} \setlength{\labelwidth}{1em}
      \setlength{\labelsep}{0.5em} } }

\newcommand{\squishend}{
    \end{list}  }
    
\newcommand{\topcaption}{%
\setlength{\abovecaptionskip}{10pt}%
\setlength{\belowcaptionskip}{0pt}%
\caption} 

\begin{document}
% \linenumbers
% \title{Multi-Task Learning with Multi-View Graph Convolutional Networks for Link Prediction and Node Classification
% }

\title{Multi-Task Representation Learning with Multi-View Graph Convolutional Networks}

\author{Hong Huang,~\IEEEmembership{Member,~IEEE,}
        Yu Song,
        Yao Wu,~\IEEEmembership{Student Member,~IEEE,}
        Jia Shi,
        Xia Xie,
        and Hai Jin,~\IEEEmembership{Fellow,~IEEE}% <-this % stops a space
\thanks{All the authors are with the National Engineering Research Center
for Big Data Technology, Service Computing Technology
and System Lab, Cluster and Grid Computing Lab and School of Computer Science and Technology, Huazhong University of Science and Technology, Wuhan, 430074, China. E-mail: \{honghuang, yusonghust, yaowu, shijia, shelicy, hjin\}@hust.edu.cn. Hong Huang is the corresponding author.}% <-this % stops a space
}

\maketitle

\begin{abstract}
Link prediction and node classification are two important downstream tasks of network representation learning. Existing methods have achieved acceptable results but they perform these two tasks separately, which requires a lot of duplication of work and ignores the correlations between tasks. Besides, conventional models suffer from the identical treatment of information of multiple views, thus they fail to learn robust representation for downstream tasks. To this end, we tackle link prediction and node classification problems simultaneously via multi-task multi-view learning in this paper. We first explain the feasibility and advantages of multi-task multi-view learning for these two tasks. Then we propose a novel model named as MT-MVGCN to perform link prediction and node classification tasks simultaneously. More specifically, we design a multi-view graph convolutional network to extract abundant information of multiple views in a network, which is shared by different tasks. We further apply two attention mechanisms: view attention mechanism and task attention mechanism to make views and tasks adjust the view fusion process. Moreover, view reconstruction can be introduced as an auxiliary task to boost the performance of the proposed model. Experiments on real-world network datasets demonstrate that our model is efficient yet effective, and outperforms advanced baselines in these two tasks.
\end{abstract}


\section{Introduction} \label{intro}

As the networks are widespread in the real world, such as academic networks~\cite{tang2008arnetminer}, biological networks~\cite{li2017scored} and social networks~\cite{van2014online,romero2016social}, learning to make predictions or classifications for networks is appealing to a wide range of concerns. Specifically, we study two fundamental tasks for network analysis in this work: link prediction and node classification. The link prediction task is defined as estimating the existence of edges between node pairs based on network observation, while the node classification task aims to assign different class labels to the nodes.

In recent years, a great deal of efforts have been devoted to solving these two tasks, such as network embedding and graph convolutional networks. Network embedding~\cite{cui2018survey} aims to learn low-dimensional representations for nodes of a network. For example, DeepWalk~\cite{perozzi2014deepwalk} adopts random walk to generate node sequences as word sequences then utilizes Skip-gram~\cite{mikolov2013distributed} model to get the node representations. Since many network embedding methods~\cite{tang2015line,wang2016structural,ribeiro2017struc2vec,grover2016node2vec} can preserve the structure and property of networks, they are very suitable for link prediction and node classification tasks. Apart from network embedding, graph convolutional networks~(GCNs)~\cite{defferrard2016convolutional,bruna2013spectral} are semi-supervised methods applicable to link prediction and node classification applications. The main idea of GCNs is to design a convolution operation for message passing between the immediate node and its neighbors, then the GCNs can be trained by a task-specific loss like Convolutional Neural Networks~(CNNs). Based on that, Kipf $\&$ Welling~\cite{kipf2016semi} has proposed a widely used convolution operation in the spatial domain that leads to significant results in various downstream tasks. Although existing methods have achieved great results, they usually suffer from one or two main inadequacies: 

(1) They are unable to utilize information encoded in multiple views while there usually exists more than one type of proximity between nodes, yielding multiple views for networks. In general, the relationships among nodes in real-world networks are sophisticated and diverse. Taking academic network as an example, we can observe the authors' interactions from three different views including co-authorship, author-citation and text-similarity views~\cite{meng2017ention}. The co-authorship view depicts the cooperation relationship between two authors, the author-citation view reflects an author citing articles by another author and the proximity of two authors in the text-similarity view is defined as the textual similarity of the papers they published. Each view may be isolated and biased, thus we need to take all views into consideration to learn a robust representation. 

\begin{figure}
    \centering
    \subfigure[\textbf{view agreement}]{\includegraphics[scale=0.47]{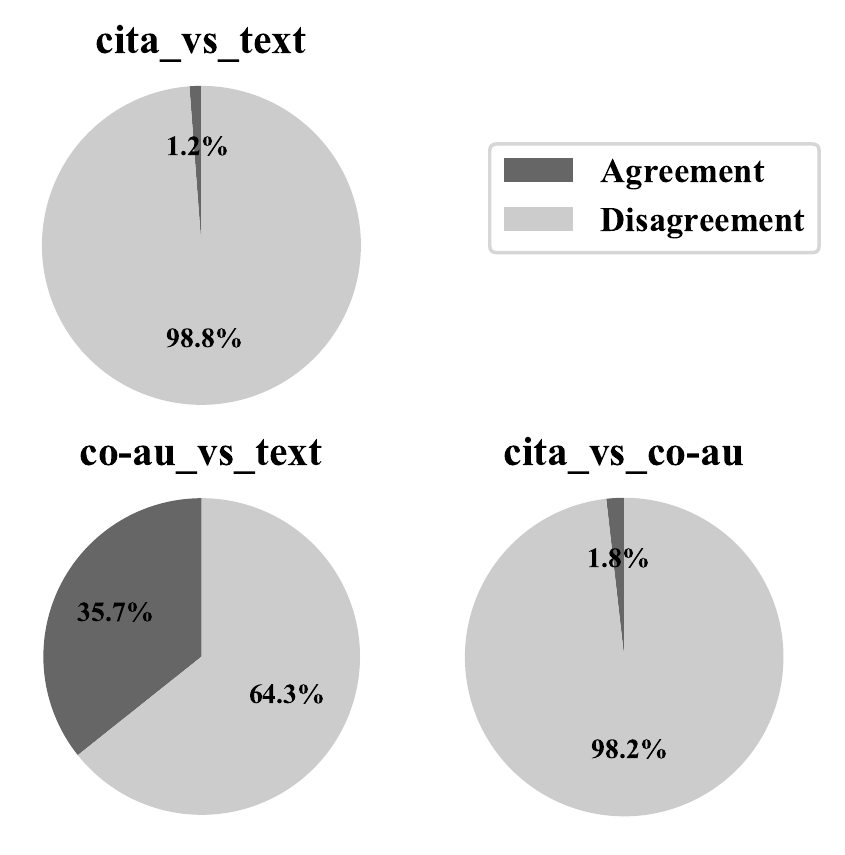}}
    \subfigure[\textbf{task correlation}]{\includegraphics[scale=0.47]{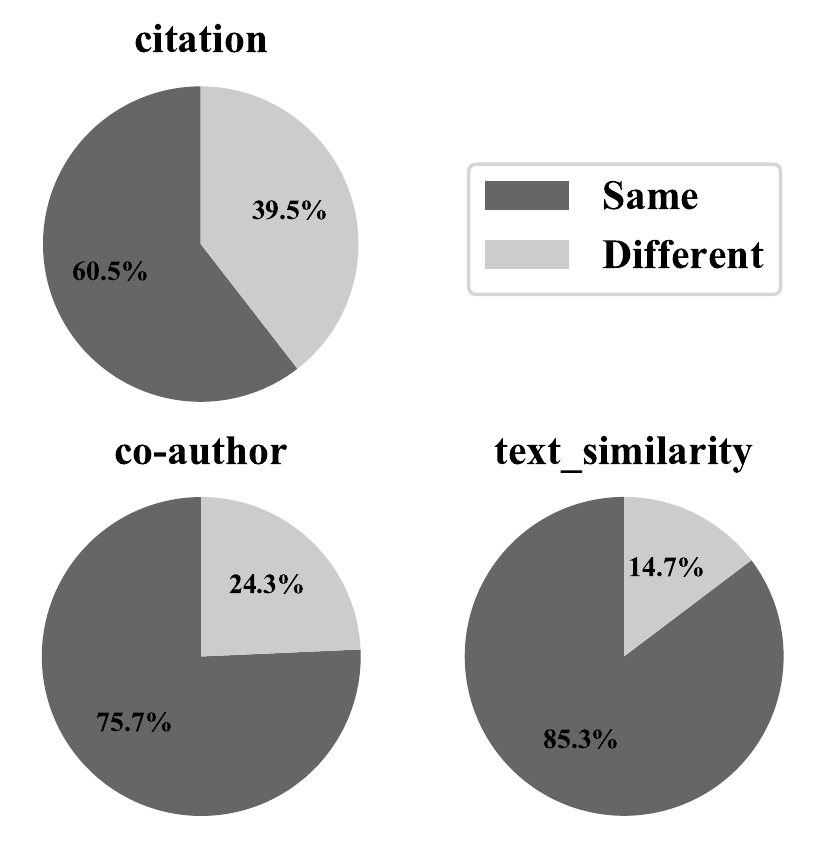}}
    \caption{(a) Case study of view agreement on AMiner Network. cita: citation network, co-au: co-authorship network and text: text similarity network. (b) Case study of task correlation on AMiner Network. Correlated represents two nodes on the edge have the same label.}
    \label{fig:case}
\end{figure}

It is of great significance to promote the collaboration of different views for learning robust representations of nodes for downstream tasks. The relationships between different views are usually complex, making the collaboration of views quite difficult. Existing multi-view models apply naive view combination strategies, such as weighted average~\cite{meng2017ention}, add~\cite{ma2018multi} and multi-view matrix factorization~\cite{zhao2017multi}, lack of sufficient collaboration of views, leading to the sub-optimal learnt representations. Besides, these methods adopt the fusion of views depending entirely on the characteristics of the data, such as view agreement or disagreement~\cite{sun2013survey}, which is not able to capture the abundant information carried by multiple views. In order to further confirm our statement, we follow the method~\cite{shi2018mvn2vec} to study the agreement level among views on AMiner network~(see details in Sec.~\ref{dataset}). Given a pair of views, each node can connect to a unique neighbor set in each view. If the Jaccard coefficient~\cite{niwattanakul2013using} between the two sets of neighbors is large than 0.5 then we think the agreement information is carried by the node in these two views, otherwise is disagreement information. Figure~\ref{fig:case}(a) shows the proportion of nodes carried agreement or disagreement information in each pair of views. As we can see, co-authorship view and text similarity view have noticeable agreement information while other pairs of views are totally different, thus it is inappropriate to simply fuse multiple views relying on agreement or disagreement.

(2) They solve these two tasks separately, which requires a large amount of repetitive work and neglects the rich correlated information between tasks. Generally, training multiple tasks simultaneously can bring more benefits than training a single task independently. First, both link prediction task and node classification task can be regarded as classification problems, with one at edge-level while the other at node-level, thus these two tasks can be optimized simultaneously. Second, these two tasks are relevant in many cases. For example, the authors in the same research filed may cite each other more often than cite other authors in different research fields; academic co-authorships are also more common among scholars in the same field and the scholars with a common research field will have higher text similarity in their papers. As shown in Figure~\ref{fig:case}(b), if there exists an edge between two nodes (authors), they will have the same label (research filed) with a high probability in each view, which indicates there is an intrinsic correlation between link prediction and node classification tasks, and multi-task learning may boost performance potentially. To this end, we are eager to design an effective model to achieve multi-task learning. It is also reasonable that different views' information contributes unequally to different tasks, thus the fusion of views needs to be decided by both the data and the targeting tasks.

To address the above limitations, we propose a novel multi-task multi-view learning framework for link prediction and node classification, namely MT-MVGCN. Specifically, our framework first takes multi-view graph convolutional networks as the backbone to learn representations of nodes in each view. After that, to encourage the collaboration between different views we design two attention mechanisms: view attention for voting the consensus representation and task-specific attention for getting specific representation for each task. Then we sum these two representations for each task and train all tasks together. By setting the multi-view graph convolutional networks and view attention shared by tasks, we allow different tasks to transfer information implicitly. Further, we add view reconstruction as an auxiliary task, which can boost the performance of our model in some scenarios. We empirically evaluate our model on four real-world datasets, and the experimental results also indicate the effectiveness of the attention mechanisms and the benefits of multi-task learning.

In summary, our main contributions are listed as follows:
\squishlist
    \item We explore the feasibility of multi-task multi-view learning for link prediction and node classification tasks, then we develop a novel framework, namely MT-MVGCN, to train these two tasks simultaneously. 
    \item Specifically, we design view attention mechanism to capture the consensus information of different views, meanwhile the task attention mechanism aims to extract the significant information for each task. Moreover, the view reconstruction task is introduced as an auxiliary task, which helps to learn more robust hidden representations.
    \item Experimental results on several multi-view networks prove our method is effective yet efficient over state-of-the-art baseline approaches. 
\squishend

The rest of this paper is organized as follows. Sec.~\ref{rework} reviews the related work. Sec.~\ref{problem} formalizes the problems; Sec.~\ref{mtmvgcn} introduces the proposed model. Sec.~\ref{exp} describes the design of experiments and reports the experimental results. Sec.~\ref{conclusion} presents the conclusion and future work. 

\section{Related Work} \label{rework}

\noindent \textbf{Network Embedding.} Our work is related to network embedding, which aims to find a nonlinear function to embed the raw network into a low-dimensional latent space.  Non-negative matrix factorization is widely used for network embedding~\cite{wang2017community,qiu2018network} because it makes all decomposed components non-negative and achieves linear dimension reduction at the same time~\cite{lee2001algorithms}. Besides, random walk is also used for various models such as DeepWalk~\cite{perozzi2014deepwalk}, node2vec~\cite{grover2016node2vec}, Struc2vec~\cite{ribeiro2017struc2vec} and SDAE~\cite{cao2016deep}. Apart from these methods, there exists some heterogeneous network embedding models. For example, Metapath2vec and HIN2Vec~\cite{dong2017metapath2vec,fu2017hin2vec} learn network embedding based meta-path for synchronized modeling of structural information and semantic association in heterogeneous networks. Shine~\cite{wang2018shine} extracts potential representations of users from heterogeneous networks and predicts unobserved signs of emotional connections.

\noindent \textbf{Graph Neural Networks.} The convolutional neural networks are designed for the data represented as a regular grid in the Euclidean space such as pictures, which is not suitable for non-euclidean data such as networks. Inspired by the success of CNNs, various works~\cite{kipf2016semi,defferrard2016convolutional,monti2017geometric,gilmer2017neural} attempt to re-define a convolutional operation on graphs. Generally, Graph Neural Networks~(GNNs) can be regraded as a sample strategy for neighbors of nodes and update the representations of nodes by a weighted sum of their neighborhoods.  Based on there methods, some improvements or variants are proposed. For example, Graph Attention Network~(GAT)~\cite{velivckovic2017graph} introduces attention mechanism to aggregate neighborhoods for nodes and FastGCN~\cite{chen2018fastgcn} aims to accelerate the speed of original GCN via importance sampling. Traditional deep neural network is a stack of non-linear layers thus it is also suitable for mapping original graph structure and properties into a low dimensional space. Some representative works, such as SDNE~\cite{wang2016structural}, SiNE~\cite{wang2017signed} and Deepcas~\cite{li2017deepcas}, provide end-to-end solutions to network problems.

\noindent \textbf{Link prediction and Node classification.} Link prediction and node classification are two of the most fundamental downstream tasks on network analysis, have attracted extensive attention from industry and academia~\cite{cui2018survey}. For example, node2vec~\cite{grover2016node2vec} performs link prediction on a social network, a biological network and an academic network. SiNE~\cite{wang2017signed} demonstrates the excellent performance of signed network embedding on link prediction. The node classification is also widely used for different networks. Many studies~\cite{meng2017ention,fu2017hin2vec,kipf2016semi,velivckovic2017graph} also apply their models for node classification, which achieves superior performance. To name a few, in language networks, such as Wikipedia, node classification can infer the Part-of-Speech tags for words~\cite{tu2016max}. In protein-protein interactions networks, node classification is applied to classify proteins into 50 different biological states~\cite{grover2016node2vec}. However, our method is the first attempt to solve these two tasks simultaneously, which can fully utilize the correlation between tasks to boost the performance.

\noindent \textbf{Multi-task learning, Multi-view  Learning and Attention Mechanisms.} Multi-task and Multi-view learning are significant fields for deep learning and great of methods have been proposed in recent years. There are many successful methods and applications in various fields~\cite{zhang2014facial,trivedi2010multiview,liu2017adversarial}, such as computer vision, data mining and natural language processing. However, only a few of works focus on networks. For multi-view networks, existing methods adopts some naive methods to learn the representations for downstream tasks. For example, MVE~\cite{meng2017ention} uses the weighted average of different views and MINES~\cite{ma2018multi} adds all views together. For multi-task learning, MTGAE~\cite{Tran-LoNGAE:2018} designs a shared auto-encoder for all tasks. Although these methods mentioned in related work are effective and efficient for their problems, they are hardly satisfactory for multi-view and multi-task learning simultaneously. However, considering multi-task and multi-view learning at the same time is a trend in recent years\cite{lu2017multilinear,zhang2016multi}. Besides, our work is related to attention mechanisms, which have been making great successes on many problems, including machine translation~\cite{vaswani2017attention}, recommendation~\cite{khattar2018hram}, image classification~\cite{mnih2014recurrent} and so on. Attention mechanisms~\cite{vaswani2017attention,meng2017ention,velivckovic2017graph} aim to learn importance of different parts of the training data so that the models can focus on the most informative parts.

\section{Problem Definition} \label{problem}

\begin{definition}\label{def1}
{\bf Multi-view network}
Given a multi-view network $G = (V, E_1, E_2, \cdots, E_k)$, where $V$ is the set of nodes  and $E_i~(1\leq i \leq k)$  is the set of edges observed from view  $i$. Each view is regarded as a single-view network reflecting a single and distinct relationship among nodes, described by edges in $E_i$.
\end{definition}

Obviously, we can utilize various methods to learn low dimension representation $\bm{Z}_i \in$ \bm{$R^d$} (d $\ll$ $|V|$) for view $i$. After that, we still face the problem of how to fuse representations of multiple views to get the final representation for multi-task learning.

\begin{problem}\label{prob1}
{\bf View Fusion}
Assuming we have obtained view representations $\{ \bm{Z}_1, \bm{Z}_2, \cdots, \bm{Z}_k \}$, the view fusion process aims to learn a fusion function $\bm{f}$ to get the final representation $ \bm{Z} = \bm{f} ( \bm{Z}_1,\bm{Z}_2,\cdots,\bm{Z}_k )$. 
\end{problem}

\begin{problem}\label{prob2}
{\bf Multi-task Learning}
Given M related or partially related tasks denoted as $\{ T_m \}_{m=1}^M$, multi-task learning aims to use the knowledge contained in all or part of the M tasks to improve the learning of model.
\end{problem}

\begin{figure*}[t]
    \centering
    \includegraphics[scale=0.45]{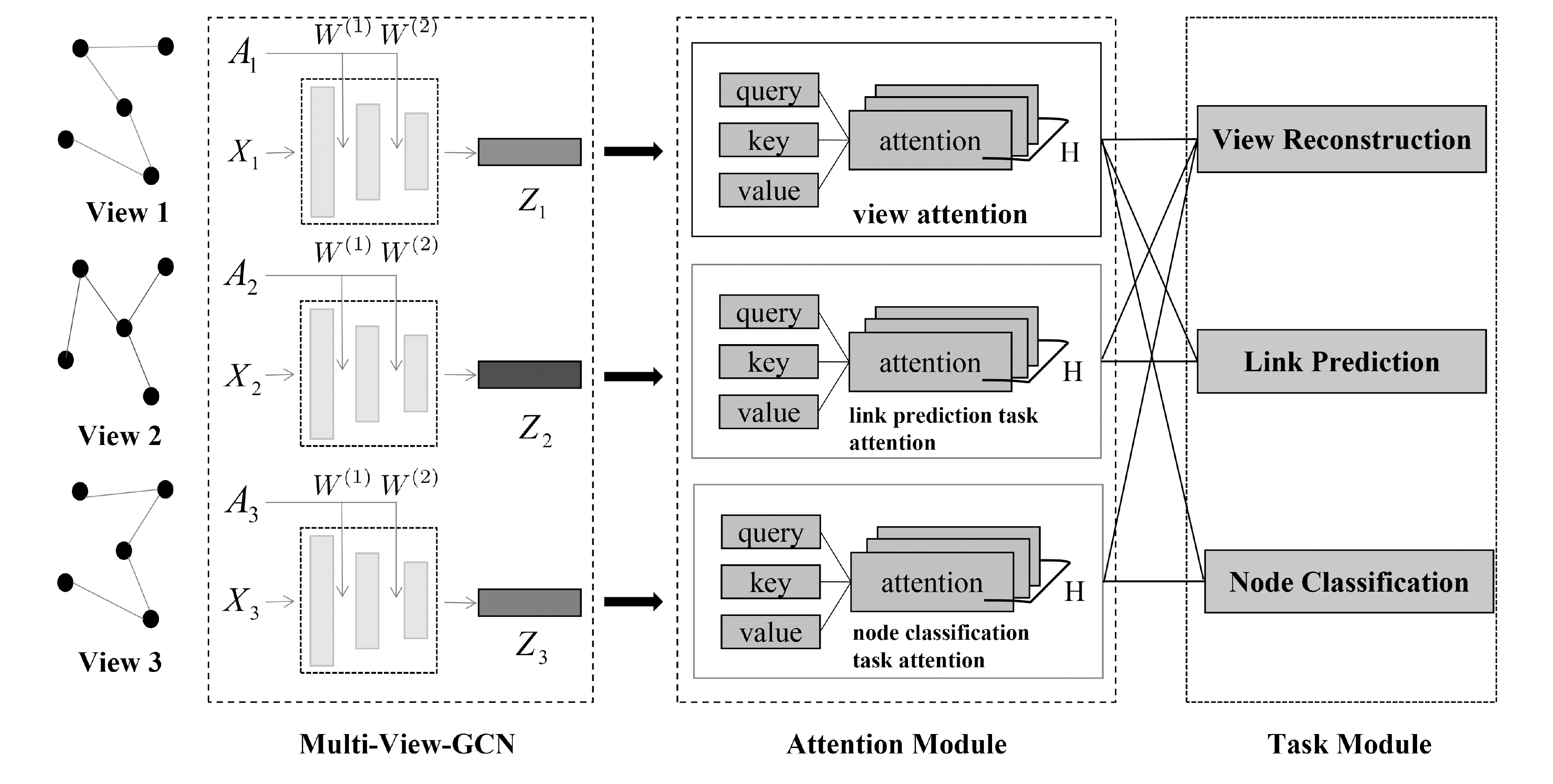}
    \caption{Overview of the proposed MT-MVGCN model. The Multi-view GCN and view attention modules are shared by different tasks. The view reconstruction module is designed as an auxiliary task.} 
    \label{fig:framework}
\end{figure*}

\section{The MT-MVGCN Framework} \label{mtmvgcn}

In this section, we will introduce our model in details. Taking three views as an example, as illustrated in Figure~\ref{fig:framework}, we first design multi-view graph convolutional networks to extract information in different views, then the view attention and task attention are applied to learn representations for link prediction and node classification tasks. Moreover, we introduce view reconstruction as an auxiliary task to improve our model. Overall, since tasks are related, it is reasonable to assume that different tasks share a common representation for the original features. Thus the multi-view graph convolutional networks and view attention mechanism are shared by tasks. Through training data in all tasks, a stronger representation can be learned for each task, and this representation can bring performance improvements. Besides, we allow different tasks to extract task-specific information through task attention mechanism.

\subsection{Multi-View Graph Convolutional Networks}
For a multi-view network $G = (V, E_1, E_2, \cdots, E_k)$, we denote adjacency matrix as $\bm{A}_i$ and feature matrix as $\bm{X}_i$ for view $i$. $N$ is the number of nodes. To extract the information encoded in view $i$, we follow Kipf $\&$ Welling~\cite{kipf2016semi} to design a Graph Convolutional Network~(GCN) with the following propagation rule at $l$-th layer:
\begin{equation}
    \bm{Z}_i^{(l)} = \sigma(\tilde{\bm{D}_i}^{-\frac{1}{2}}\tilde{\bm{A}_i}\tilde{\bm{D}_i}^{-\frac{1}{2}}\bm{Z}_i^{(l-1)}\bm{W}^{(l)}) \label{gcn}
\end{equation}
Here, we set $\tilde{\bm{A}_i} = \bm{A}_i + \bm{I}_N$ and $\bm{I}_N$ is identity matrix, $\tilde{\bm{D}}_{ii} = \sum_j\tilde{\bm{A}}_{ij}$, $\bm{W}^{(l)}$ is the weight matrix, $\bm{Z}_i^{(l)} \in \bm{R^{N\times d^{(l)}}}$, $\bm{Z}_i^{(0)} = \bm{X}_i$ and $\sigma (\cdot)$ is the activation function. In this paper, we choose $tanh$ as activation function in all cases. If the node features are not available the $\bm{X}_i$ will be an identity matrix.  

Note that there is a key point in multi-view GCN that we make the weight matrix $\bm{W}^{(l)}$ shared by each view. Through this shared architecture\footnote{Note the node set is shared across all views. Moreover, the multi-view GCN can also be applied to such multi-view networks that have a few nodes unobserved in each view. In this case, we just need to pad those unobserved nodes' entries as zero in the adjacency matrix of each view.} we can: (1) project all views into the same semantic space so that the fusion of view representations is more interpretable. (2) make our model scalable for views and take up less memory due to the parameters shared mechanism. (3) allow different views influence each other  mutually and collaborate implicitly. After getting the final output of multi-view GCN for each view denoted as $\{\bm{Z}_1,\bm{Z}_2,\cdots,\bm{Z}_k \}$, we will discuss how to design the fusion function $f$ to generate representation for each task in the next section.

\begin{figure}[t]
    \centering
    \includegraphics[scale=0.34]{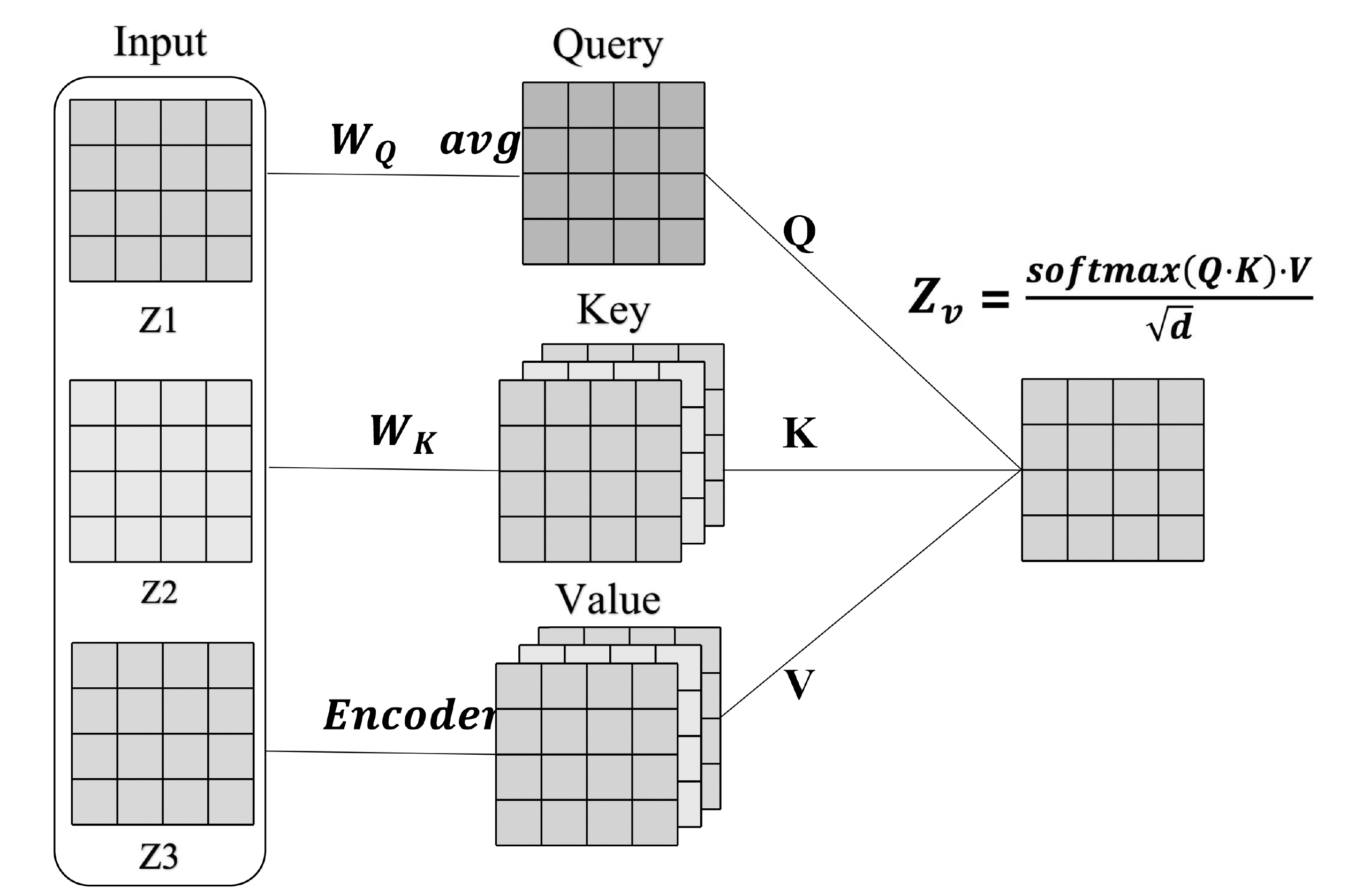}
    \caption{Illustration of View Attention Mechanism. }
    \label{fig:att}
\end{figure}

\subsection{Attention based View Fusion}

In order to achieve the collaboration between different views, we need to fuse view-specific representations by a flexible way. Instead of concatenating or adding the representations in different views, we can also apply attention mechanism to capture the complex relationships between views and tasks. Concretely, view attention mechanism is proposed to make all views vote the robust representation and task attention mechanism aims to fetch the important information across different views for each task.

\subsubsection{View Attention Mechanism}
Recent research shows that the attention mechanism can be understood as the weighted sum of values, and the query and keys are used to calculate the weight coefficients of the corresponding values. In this work, we still apply the query, keys to calculate the attention score for each view, then we sum view-specific values according to the attention score. However, the scale of the networks are usually relative large thus we must design a time efficient way to calculate attention score for each view. Here we first introduce view attention mechanism and take three views as an example, which is illustrated in Figure~\ref{fig:att}.

As an initial step, weight matrices $\bm{W}_Q$ and $\bm{W}_K$ define two learnable linear transformations for the node representations of each view respectively, then we get the query vector $\bm{Q}$ and key vectors $\bm{K}$ as follows:

\begin{equation}
    % Q = avg[Z_1W_Q^1,Z_2W_Q^2,Z_3W_Q^3] = avg[Q_1,Q_2,Q_3]
    \bm{Q}_i = \bm{Z}_i \bm{W}_Q, \quad \bm{Q} = avg[\bm{Q}_1,\bm{Q}_2,\bm{Q}_3]
\end{equation}

\begin{equation}
    % K = [Z_1W_K^1,Z_2W_K^2,Z_3W_K^3] = [K_1,K_2,K_3]
    \bm{K}_i = \bm{Z}_i \bm{W}_K, \quad \bm{K} = stack[\bm{K}_1,\bm{K}_2,\bm{K}_3]
\end{equation}

\noindent where the $avg$ is element-wise average, $stack$ operation stacks all key matrices into a new matrix with rank one higher than each key matrix, $\bm{W}_Q$ and $\bm{W}_K$ are shared by all views in order to make our model be scalable for any number of views.

% For value vectors, matrix multiplication will lead to great changes in the original structural information. Thus we design a graph encoder to calculate the values in order to preserve the structural information as much as possible. The graph encoder is defined by function $g$ and value weight matrix $W_V$:

Since the output of view attention is the weighted average of value matrices, thus it is of great significance to require the value matrices to preserve original structural information as much as possible. Different from query and key matrices, we design a graph encoder to calculate the value matrices, aiming to further take the higher-order structural information into consideration. The graph encoder is defined by function $g$ and weight matrix $\bm{W}_V$:   

\begin{equation}
    \bm{V}_i = g(\tilde{\bm{A}}_i,\bm{Z}_i) = \tilde{\bm{A}}_i\bm{Z}_i\bm{W}_V,
    \quad \bm{V} = stack[\bm{V}_1,\bm{V}_2,\bm{V}_3] \label{gae}
\end{equation}

After getting query, key, value matrices, we calculate the attention score with the intuition: the query matrix is the average information of views thus it can represent the consensus information of views. If the query matrix $\bm{Q}$ and the key matrix $\bm{K}_i$ have a larger inner product through broadcasting, meaning all views believe view $i$ is more informative, then we will assign a larger weight for this view. Following such intuition, we compute the final representation of view attention mechanism as:
\begin{equation}
    \bm{Z}_v = \frac{softmax(\bm{Q}\cdot \bm{K})\cdot \bm{V}}{\sqrt{d}} \label{att_score}
\end{equation}
where the $softmax$ function is used to normalize all choices, $d$ is the hidden dimension of these matrices. In practice, we may not compute the attention score with matrix multiplication due to the node size limitation in real-world networks. Compared with matrix multiplication, the inner product operation reduces the time complexity by $N$ times, which greatly accelerates the running speed.

% Specially, we extend Equ. ~\ref{att_score} to apply multi-head attention\cite{vaswani2017attention} to learn attention in different subspace. H attention mechanisms are computed independently and then the results will be concatenated to get the following final representation:

Since multiple views can be significantly different from each other, the variance of network data are quite high. To tackle such problem, we extend the Eq.~\eqref{att_score} to multi-head attention to stabilize the learning process of view attention, similarly to ~\cite{vaswani2017attention}. Specifically, we utilize $H$ attention heads to execute Eq.~\eqref{att_score} independently and simultaneously, and then their results are concatenated, leading to the following output feature representation:  
\begin{equation}
    \bm{Z}_v = \lVert_{h=1}^H (\frac{softmax(\bm{Q}^h\cdot \bm{K}^h)\cdot \bm{V}^h}{\sqrt{d^h}}) \label{mulhead}
\end{equation}
where $\lVert$ represents concatenation, and $h$ means the $h$-th head.

\subsubsection{Task Attention Mechanism}

Similar to the view attention mechanism, task attention mechanism is also computed according to Eq. ~\eqref{t_mulhead}. 
\begin{equation}
\begin{aligned}
    & \bm{Z}_t = \lVert_{h=1}^H (\frac{softmax(\bm{Q}_t^h\cdot \bm{K}_t^h)\cdot \bm{V}^h}{\sqrt{d^h}})
\end{aligned}
\label{t_mulhead}
\end{equation}
Where the $\bm{V}$ is also shared with view attention mechanism, but the main difference between Eq.~\eqref{mulhead} and Eq.~\eqref{t_mulhead} is that the $\bm{Q}_t$ and $\bm{K}_t$ are replaced by two trainable variables for each task. Due to $\bm{Q}_t$ and $\bm{K}_t$ do not know the information of views, they are trained totally by the specific task so that the task can gather useful information through optimizing these two variables. In other words, this strategy enables different tasks to assign customized weights to different views, thus only the information is believed serviceable will be chosen by each task.  Besides, we should notice that view attention result can be shared by all tasks and task attention results are unique across tasks. 

After we compute the results of view attention mechanism and task attention mechanism for $m$-th task, denoted as $\bm{Z}_v^m$ and $\bm{Z}_t^m$ respectively. We integrate the results by a hyper parameter $\alpha$ ranging from $0 \sim 1$ to obtain the final representation:

\begin{equation}
    \bm{Z}^m = \alpha \bm{Z}_t^m + (1 - \alpha) \bm{Z}_v^m
\end{equation}

\subsection{Multi-task Learning}

In this work, we study two significant tasks associated with relational: link prediction and node classification. For link prediction task, cosine similarity implies the distance between the pair of nodes in vector space, thus the cosine similarities between node pairs are estimated as features to make predictions. Given an edge $e$, we predict the existence of $e$ in each view; $y$ is the label of edge $e$, in which the $k$-th dimension is set as 1 if $e$ exists in $k$-th view else set as 0, which comes down to a multi-label classification problem. As there exists a huge amount of node pairs that share no edges, predicting edge existence between all possible node pairs is unrealistic. Therefore, we only predict the existence of edges that appear in at least one view. For node classification task, we use the learnt representations as features and the node labels are already provided. For $m$-th task, $\bm{Z}^m$ is the final feature representation, $y^m$ is the ground truth label, we first predict the label using a nonlinear layer: $p^m = f(\bm{Z}^m\bm{W}^m + b^m)$, $f$ is activation function (eg. sigmoid or softmax function), $\bm{W}^m$ is the weight matrix and $b^m$ is bias. The cross-entropy is applied for prediction loss, thus the overall loss is formulated as:

\begin{equation}
    L = -\sum\limits_{m=1}^M{\lambda_m \sum\limits_{j=1}^{N_b} [y_j^m\log{p_j^m} + (1-y_j^m)\log{(1-p_j^m)}]}
\end{equation}

\noindent where $\lambda_m$ is a hyper-parameter determines the importance of $m$-th task, $M$ is the number of tasks, and $N_b$ is the batch size.

\subsection{MT-MVGCN++}

The attention mechanisms in MT-MVGCN employ graph encoder to preserve the structural information when computing the value vectors in Eq.~\eqref{gae}. However, the structural information can be preserved mainly because the adjacency matrix contains abundant information rather than the original information is preserved. In addition, graph encoder may lead the features to be over smoothed~\cite{li2018deeper}. Here, we introduce an auxiliary task: view reconstruction, which aims to decode the values back to the original feature space to constrain the encoder preserving enough original features. As \cite{salakhutdinov2009semantic} proved, the reconstruction criterion can smoothly capture the data manifolds thus preserve the original information. 
More specifically, we utilize the final representations of tasks to reconstruct a representation of each view: $\widetilde{\bm{Z}}_i = \tilde{g}(\tilde{\bm{A}}_i,\bm{Z}^{\prime}) = \tilde{\bm{A}}_i \bm{Z}^{\prime} \bm{W}_D^i$, where $\bm{Z}^{\prime} = \sum_{m=1}^M \bm{Z}^m$.

In this case, view reconstruction can be regarded as an unsupervised task to be trained with other tasks together. The objective function of view reconstruction is mean square error:

\begin{equation}
    loss = \frac{1}{k}\frac{1}{N}\sum\limits_{i=1}^k{\Vert \bm{Z}_i-\widetilde{\bm{Z}}_i \Vert}_F^2
\end{equation}

\noindent where $k$ is the number of views, $N$ is the node size. To this end, the objective function is the sum of prediction loss and reconstruction loss:

\begin{equation}
\begin{aligned}
    L &= -\sum\limits_{m=1}^M{\lambda_m \sum\limits_{j=1}^{N_b} [y_j^m\log{p_j^m} + (1-y_j^m)\log{(1-p_j^m)}]} \\ &+ \lambda_{M+1} \frac{1}{k}\frac{1}{N}\sum\limits_{i=1}^k{\Vert \bm{Z}_i-\widetilde{\bm{Z}}_i \Vert}_F^2
\end{aligned}
\end{equation}

\subsection{Implementation}
In practice, we utilize Tensorflow~\cite{abadi2016tensorflow} for an efficient GPU-based implementation of the proposed models, then the parameters can be optimized efficiently and automatically with gradient descent and back propagation algorithm. Due to the sparsity of network data, we use sparse-dense matrix multiplication for Eq.~\eqref{gcn}, as described in \cite{kipf2016semi}. Through stacking multiple GCN layers, as described in Eq.~\eqref{gcn}, we construct a multi-view GCN module to extract the information in each view. Then the outputs of multi-view GCN, i.e. $\{\bm{Z}_1,\bm{Z}_2,\cdots,\bm{Z}_k \}$, will be fused by view attention module and task attention module, and yield the hidden representation $\bm{Z}^m$ for $m$-th task. After that, we perform multiple tasks simultaneously to calculate the overall loss. Finally, the parameters can be updated through minimizing the overall loss. The reference code is available at \url{https://github.com/yusonghust/MT-MVGCN}

\section{Experiment} \label{exp}

\begin{table}[htbp]
\centering
\setlength{\tabcolsep}{1mm}
\renewcommand\arraystretch{1.2}
\caption{Statistics of the datasets}
\label{datasets}
\begin{tabular}{c|c|c|c|c|c}
\hline
\textbf{Dataset}                           & \textbf{\#  views}           & \textbf{\# nodes}                & \textbf{\# edges} & \textbf{\# labels}                   & \textbf{Tasks}               \\ \hline \hline
\multirow{2}{*}{\textbf{YouTube}} & \multirow{2}{*}{4} & \multirow{2}{*}{5,108}  & \multirow{2}{*}{3,263,045} &\multirow{2}{*}{/} & link prediction     \\
                                  &                    &                        &                            &                   & reconstruction      \\ \hline
\multirow{2}{*}{\textbf{Twitter}} & \multirow{2}{*}{4} & \multirow{2}{*}{12,741} & \multirow{2}{*}{3,154,719} &\multirow{2}{*}{/} & link prediction     \\
                                  &                    &                        &                            &                    & reconstruction      \\ \hline
\multirow{3}{*}{\textbf{Flickr}}  & \multirow{3}{*}{2} & \multirow{3}{*}{34,881} & \multirow{3}{*}{3,290,030} &\multirow{3}{*}{169} & link prediction     \\
                                  &                    &                        &                            &                    & node classification \\
                                  &                    &                        &                            &                    & reconstruction      \\ \hline
\multirow{3}{*}{\textbf{Aminer}}  & \multirow{3}{*}{3} & \multirow{3}{*}{8,438}  & \multirow{3}{*}{2,433,356} &\multirow{3}{*}{8}  & link prediction     \\
                                  &                    &                        &                            &                    & node classification \\
                                  &                    &                        &                            &                    & reconstruction      \\ \hline
\end{tabular}
\end{table}

\begin{table}[htbp]
\centering
\renewcommand\arraystretch{1.2}
\topcaption{Configurations of the proposed models}
\label{cfg}
\begin{tabular}{c|c|c|c|c}
\hline
\textbf{Datasets}       & \textbf{YouTube} & \textbf{Twitter} & \textbf{Flickr} & \textbf{AMiner} \\ \hline \hline
GCN-Layers              & 3                & 3                & 2               & 1               \\ \hline
GCN-Hiddensize          & 64               & 64               & 32              & 32              \\ \hline
$\alpha$                & 0.5              & 0.5              & 0.5             & 0.5             \\ \hline
$\lambda_{linkpred}$    & 1.0              & 1.0              & 1.0             & 1.0             \\ \hline
$\lambda_{nodecls}$     & -                & -                & 0.1           & 0.001             \\ \hline
$\lambda_{reconstruction}$  & 0.01         & 0.01         & 0.01        & 0.01            \\ \hline
\end{tabular}
\end{table}

\begin{table*}[htbp]
\centering
\setlength{\tabcolsep}{4mm}
\renewcommand\arraystretch{1.2}
\topcaption{Quantitative results on the link prediction task for different datasets. }
\label{linkpred}
\begin{tabular}{c|c|cc|cc|cc|cc}
\hline
\multirow{2}{*}{\textbf{Category}} & \multirow{2}{*}{\textbf{Methods}} & \multicolumn{2}{c|}{\textbf{YouTube}} & \multicolumn{2}{c|}{\textbf{Twitter}} & \multicolumn{2}{c|}{\textbf{Flickr}} & \multicolumn{2}{c}{\textbf{AMiner}} \\ \cline{3-10} 
                                   &                                   & \textbf{AP}      & \textbf{AUC}      & \textbf{AP}      & \textbf{AUC}      & \textbf{AP}      & \textbf{AUC}     & \textbf{AP}      & \textbf{AUC}     \\ \hline \hline
\multirow{2}{*}{STSV}              & DeepWalk                          & 0.697             & 0.617             & 0.631             & 0.717            & 0.737             & 0.740            & 0.732             & 0.764            \\
                                   & GCN                               & 0.708             & 0.779             & 0.629             & 0.566             & 0.759             & 0.768            & 0.789             & 0.831            \\ \hline
\multirow{3}{*}{STMV}              & DW-con                            & 0.701             & 0.719             & 0.633             & 0.566             & 0.707             & 0.716            & 0.751             & 0.775            \\
                                   & MVE                               & 0.723             & 0.627             & 0.642             & 0.546             & 0.669             & 0.676            & 0.709             & 0.736            \\
                                   & HIN2Vec                           & 0.798             & 0.776             & 0.653             & 0.625             & 0.802             & 0.818            & 0.626             & 0.636            \\ \hline
MTSV                               & MTGAE                             & 0.782             & 0.798             & 0.649             & 0.619             & 0.783             & 0.808            & 0.786             & 0.833            \\ \hline
\multirow{3}{*}{MTMV}              & GCN-con                           & 0.786             & 0.775             & 0.661             & 0.710             & 0.824             & 0.927            & 0.816             & \textbf{0.857}   \\
                                   & MT-MVGCN                          & 0.800             & 0.804             & \textbf{0.669}    & 0.703             & \textbf{0.847}    & \textbf{0.955}   & \textbf{0.837}    & \textbf{0.858}   \\
                                   & MT-MVGCN++                        & \textbf{0.807}    & \textbf{0.824}    & 0.672             & \textbf{0.719}    & \textbf{0.861}    & 0.949            & 0.829             & 0.844            \\ \hline
\end{tabular}
\end{table*}

% \begin{table}[htbp]
% \centering
% \setlength{\tabcolsep}{4mm}
% \renewcommand\arraystretch{1.2}
% \topcaption{Statistics of the datasets}
% \label{datasets}
% \begin{tabular}{c|c|c|c|c}
% \hline
% \textbf{Datasets} & \textbf{\# Nodes} & \textbf{\# Views} & \textbf{\# Edges} & \textbf{Tasks} \\ \hline \hline
% \textbf{YouTube}  & 5,108           & 4              & 3,263,045    &              \\ \hline
% \textbf{Twitter}  & 12,741          & 4              & 3,154,719    &             \\ \hline
% \textbf{Flickr}   & 34,881          & 2              & 3,290,030    &               \\ \hline
% \textbf{AMiner}   & 8,438           & 3              & 2,433,356    &             \\ \hline
% \end{tabular}
% \end{table}

\subsection{Datasets} \label{dataset}
We select four real-world multi-view network datasets. Link prediction will be evaluated in all datasets and node classification will be evaluated in the last two datasets. 

\squishlist
\item \textbf{YouTube Dataset}
For YouTube network~\cite{zafarani2009social}, four views are constructed, representing friendship, number of common friends, number of common subscribers between two users and number of common favorite videos respectively. 

\item \textbf{Twitter Dataset}
The twitter network~\cite{de2013anatomy} contains four views, including re-tweeting, reply, mention and friendship. We apply the similar method~\cite{tang2015line} to reconstruct these three views to make them denser due to the sparsity of original network. 

\item \textbf{Flickr Dataset}
The network built from Flickr dataset~\cite{tang2009relational} includes two views. Friendship view is the contact network among the blog owners. Tag-similarity view is a network with each node connecting to its top ten nearest neighbors, and the similarity is calculated based on the user's tags. The community memberships are used as node labels.

\item \textbf{AMiner Dataset}
The AMiner network~\cite{tang2008arnetminer} has three views: author-citation, co-authorship and text similarity. The edge weight in co-authorship view is the number of publications cooperated by each pair of authors; the edge weight in citation view is defined as the number of literature published by one author and cited by another author; the text similarity view depicts top ten nearest neighbors for each node, and the similarity is calculated based on the title and abstract by TF-IDF~\cite{ramos2003using}. We only preserve authors in eight research fields as~\cite{dong2017metapath2vec} as nodes and research fields are treated as node labels. 

\squishend

The details of datasets are listed in Table ~\ref{datasets}.

\subsection{Baselines}
We choose four types of baselines: single-task single-view(STSV) based, single-task multi-view(STMV) based, multi-task single-view based(MTSV) and multi-task multi-view(MTMV) based. 

\squishlist

\item \textbf{DeepWalk~\cite{perozzi2014deepwalk}}  DeepWalk uses random walk to construct node sequences then applies skip-gram model to learn network representation.
\item \textbf{GCN~\cite{kipf2016semi}} GCN is a semi-supervised method based on graph convolutions.
\item \textbf{GAT~\cite{velivckovic2017graph}} GAT introduces attentions mechanism for graph convolutional operations.
\item \textbf{DW-con}  DW-con concatenates representations learned by DeepWalk of all views to generate the final representation.
\item \textbf{MVE~\cite{meng2017ention}} MVE adopts attention mechanism to assign weights for the weighted sum of view representations.
\item \textbf{RGCN~\cite{schlichtkrull2018modeling}} RGCN allows multiple relationships among two nodes of a graph. Views corresponding to distinct relationships are encoded differently.
\item \textbf{HIN2Vec~\cite{fu2017hin2vec}} Multi-view network is a type of heterogeneous network thus HIN2Vec is adopted to learn network representation.
\item \textbf{MTGAE~\cite{Tran-LoNGAE:2018}} Multi-task graph auto-encoder defines a set of nonlinear transformations to capture graph structure for multi-task learning.
\item \textbf{GCN-con} A variant of MT-MVGCN, which concatenates the results of multi-view GCN to get the final representation.
\item \textbf{MT-MVGCN} The proposed model for multi-task learning with information of multiple views.
\item \textbf{MT-MVGCN++} A variant of MT-MVGCN, which introduces an auxiliary task: view reconstruction.

\squishend

Note that we study the attention mechanisms in section~\ref{att_case}, so here we mainly compare our models with some representative baselines. Besides, there are some multi-view matrix factorization methods, which will not be compared with our model since they can not be extended to large-scale networks.

% \usepackage{multirow}
\begin{table*}[htbp]
\centering
\setlength{\tabcolsep}{1.5mm}
\renewcommand\arraystretch{1.2}
\topcaption{Quantitative results in terms of classification \textbf{accuracy/precision/f1} scores on different datasets}
\label{nodecls}
\begin{tabular}{c|c|ccc|ccc}
\hline
\multirow{2}{*}{\textbf{Category}} & \multirow{2}{*}{\textbf{Methods}} & \multicolumn{3}{c|}{\textbf{Flickr}}             & \multicolumn{3}{c}{\textbf{Aminer}}             \\ \cline{3-8} 
                                   &                                   & \textbf{30\%}   & \textbf{50\%}   & \textbf{70\%}   & \textbf{30\%}   & \textbf{50\%}   & \textbf{70\%}   \\ \hline \hline
\multirow{3}{*}{STSV}              & DeepWalk                          & 0.898/0.668/0.699  & 0.901/0.755/0.731 & 0.913/0.738/0.723 & 0.932/0.899/0.765 & 0.936/0.912/0.899  & 0.944/0.913/0.921          \\
                                   & GCN                               & 0.907/0.690/0.720          & 0.909/0.772/0.749          & 0.911/0.737/0.723          & 0.876/0.810/0.830          & 0.879/0.843/0.821          & 0.883/0.812/0.827          \\
                                   & GAT                               & 0.901/0.675/0.706          & 0.913/0.780/0.758          & 0.918/0.752/0.738          & 0.889/0.816/0.835          & 0.899/0.871/0.852          & 0.907/0.845/0.858              \\ \hline
\multirow{4}{*}{STMV}              & DW-con                            & 0.883/0.633/0.666          & 0.883/0.719/0.693          & 0.884/0.673/0.656          & 0.875/0.806/0.826          & 0.876/0.844/0.823          & 0.877/0.805/0.820          \\
                                   & MVE                               & 0.882/0.631/0.664          & 0.891/0.734/0.710          & 0.893/0.684/0.668          & 0.878/0.800/0.821          & 0.880/0.850/0.829          & 0.879/0.809/0.824          \\
                                   & RGCN                              & 0.911/0.701/0.730          & 0.917/0.789/0.768          & 0.922/0.780/0.767          & 0.947/0.906/0.917          & 0.957/0.940/0.931          & 0.960/0.930/0.936              \\
                                   & HIN2Vec                           & 0.910/0.698/0.727          & 0.911/0.776/0.754          & 0.914/0.766/0.753          & 0.961/0.934/0.941          & 0.962/0.951/0.943          & 0.961/0.935/0.941          \\ \hline
MTSV                               & MTGAE                             & 0.895/0.661/0.692          & 0.898/0.749/0.725          & 0.903/0.712/0.696          & 0.854/0.768/0.791          & 0.865/0.825/0.802          & 0.875/0.810/0.825          \\ \hline
\multirow{3}{*}{MTMV}              & GCN-con                           & 0.896/0.663/0.695          & 0.906/0.765/0.743          & 0.914/0.750/0.736          & 0.875/0.811/0.830          & 0.875/0.847/0.826          & 0.876/0.811/0.826          \\
                                   & MT-MVGCN                          & 0.906/0.688/0.718          & \textbf{0.922/0.800/0.780} & 0.923/0.778/0.765          & 0.969/0.948/0.954          & 0.971/0.961/0.954          & 0.973/0.953/0.958          \\
                                   & MT-MVGCN++                        & \textbf{0.917/0.716/0.745} & \textbf{0.921/0.798/0.777} & \textbf{0.935/0.800/0.788} & \textbf{0.973/0.952/0.958} & \textbf{0.977/0.970/0.965} & \textbf{0.984/0.974/0.976} \\ \hline
\end{tabular}
\end{table*}

\subsection{Experimental Setup}

For the proposed model, the parameter settings are listed in Table ~\ref{cfg} for different datasets. For DeepWalk, the number and the length of random walks for each node is set as 10 and 80 respectively. The window size of the skip-gram model is 10. For RGCN, HIN2Vec, MTGAE and MVE we use the default parameter settings in the original paper. For all baselines, we set the hidden dimension of representation same as our model. We train all models using Adam~\cite{kingma2014adam} optimizer with learning rate of 0.001. We also use dropout with drop rate of 0.5 and early stopping to prevent over-fitting. For two attention mechanisms and GAT, we set the number of heads as 4. We calculate the average precision score~(AP), area under curve score~(AUC) for link prediction and accuracy score, precision score and f1 score for node classification. As graph data has a large variance, we follow the method\cite{kipf2016semi} to repeat each method for ten times, and the average metrics are reported.  For single-view based methods, the best result among different views is reported. As node features are not available for our datasets the feature matrix  will be an identity matrix. 

\subsection{Experimental Results}

In this section, the experimental results of link prediction and node classification are presented. For link prediction the percentage of training data is 50\%, and for node classification, the percentage of training data varies from 30\% to 70\%. The link prediction results are listed in Table ~\ref{linkpred} and  node classification results are listed in Table ~\ref{nodecls}. From the experimental results, we have the following observations and analysis:

\squishlist
\item For link prediction task, MTSV based methods perform better than STSV based method on Flickr and AMiner datasets. In contrast, MTSV based method has similar results with STSV based methods on YouTube and Twitter datasets. This phenomenon can prove that multi-task learning is applicable and useful in many scenarios. First, as the node labels are not available for YouTube and Twitter datasets, the MTGAE model is only trained by link prediction task, thus it is not surprising that MTGAE is close to the performance of the STSV based methods. Second, link prediction task is co-training with node classification task on Flickr and AMiner datasets. Thus we can see that multi-task learning boosts the performance on these two datasets.
\item For link prediction task, the results of MT-MVGCN++ are more satisfactory than MT-MVGCN on YouTube and Twitter datasets while the opposite is true on the other two datasets. On the one hand, this result is due to the fact that we introduce an auxiliary task for the first two datasets, which can also prove that multi-task learning is reasonable and useful. On the other hand, for the last two datasets, link prediction task is already co-trained with node classification task, thus the auxiliary task even makes the performance degrade slightly since it may introduce extra noises for the link prediction task.
\item For node classification task, we can see that multi-task based baselines have no obvious improvement but HIN2Vec and RGCN outperform single-view based baselines. However, our methods always maintains significantly performance improvement with training percentage varying from 30\% to 70\%. 
\item As we can see, the average performance of STMV based methods is close to STSV based but the performance of MTMV based methods is better than MTSV based method. The reason may be that multi-task learning can make better use of information of multiple views. Besides, we also find that some naive view fusion methods, such as concatenation~(DW-con) and average~(MVE), fail to  make collaboration the between different views, thus these methods achieve relatively poor performance. Lastly, the meta-path based heterogeneous network representation learning method has achieved satisfactory results but still not as good as our proposed method. Since there is only one type of node in the multi-view network, leading to the semantic information of the meta-path greatly reduced.
\item For both tasks, the proposed methods outperform all compared baselines on all datasets which indicates our methods can be better extend to link prediction and node classification tasks. We can find that although GCN-con is based on multi-task multi-view learning, the performance is still under our methods because our models adopt two attention mechanisms to make use of information from different views rather than combining all views naively. More specifically, view attention mechanism and task attention mechanism complement each other to better extract the information in each view. Besides, we observe that by simultaneously training view reconstruction task with MT-MVGCN, MT-MVGCN++ achieves better performance in most instances.
\squishend

\begin{figure}[t]
    \centering
    \subfigure[\textbf{YouTube and Twitter}]
    {
    \includegraphics[scale=0.32]{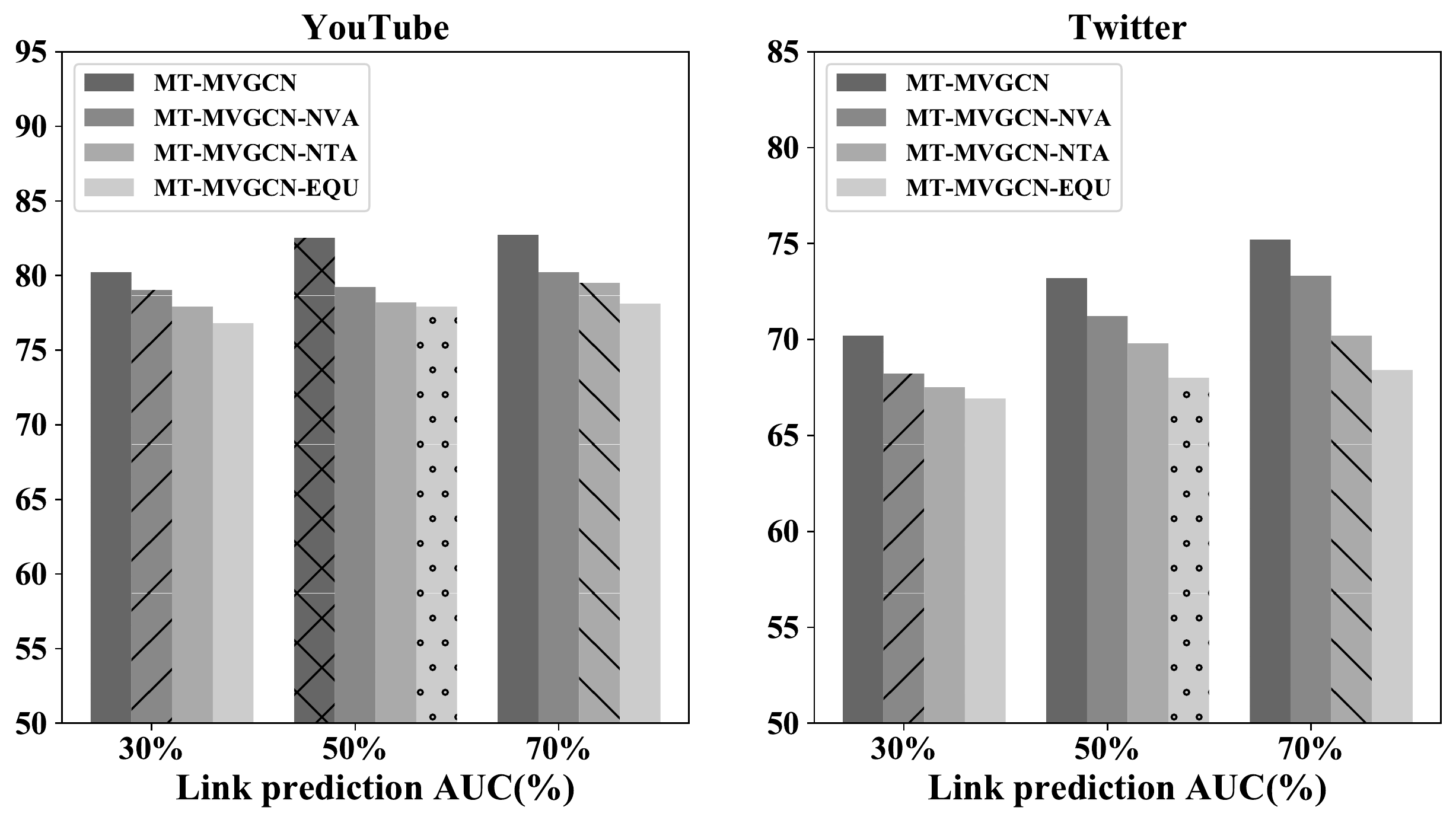}
    }
    \\
    \subfigure[\textbf{Flickr}]
    {
    \includegraphics[scale=0.32]{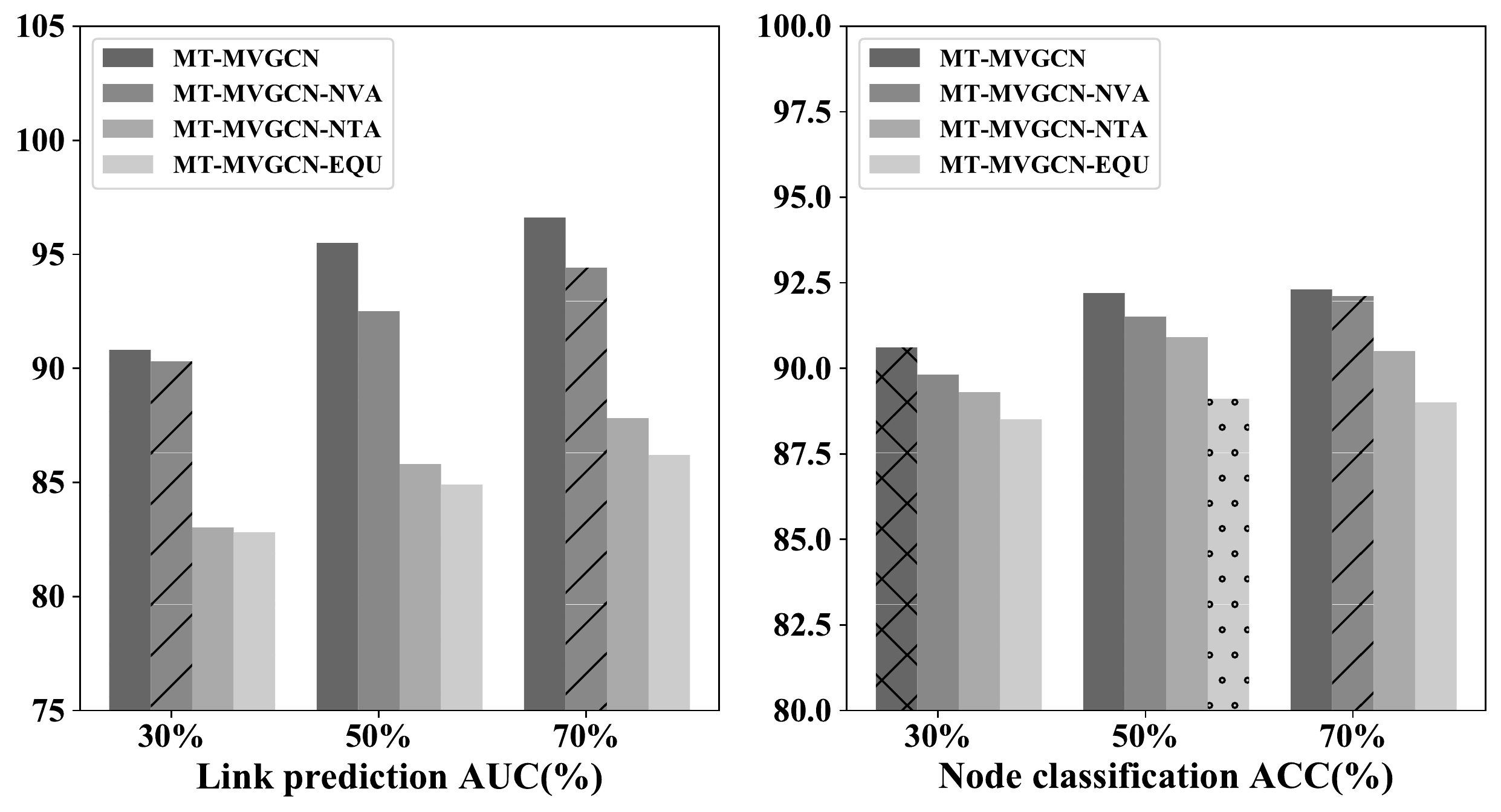}
    }
    \\
    \subfigure[\textbf{AMiner}]
    {
    \includegraphics[scale=0.32]{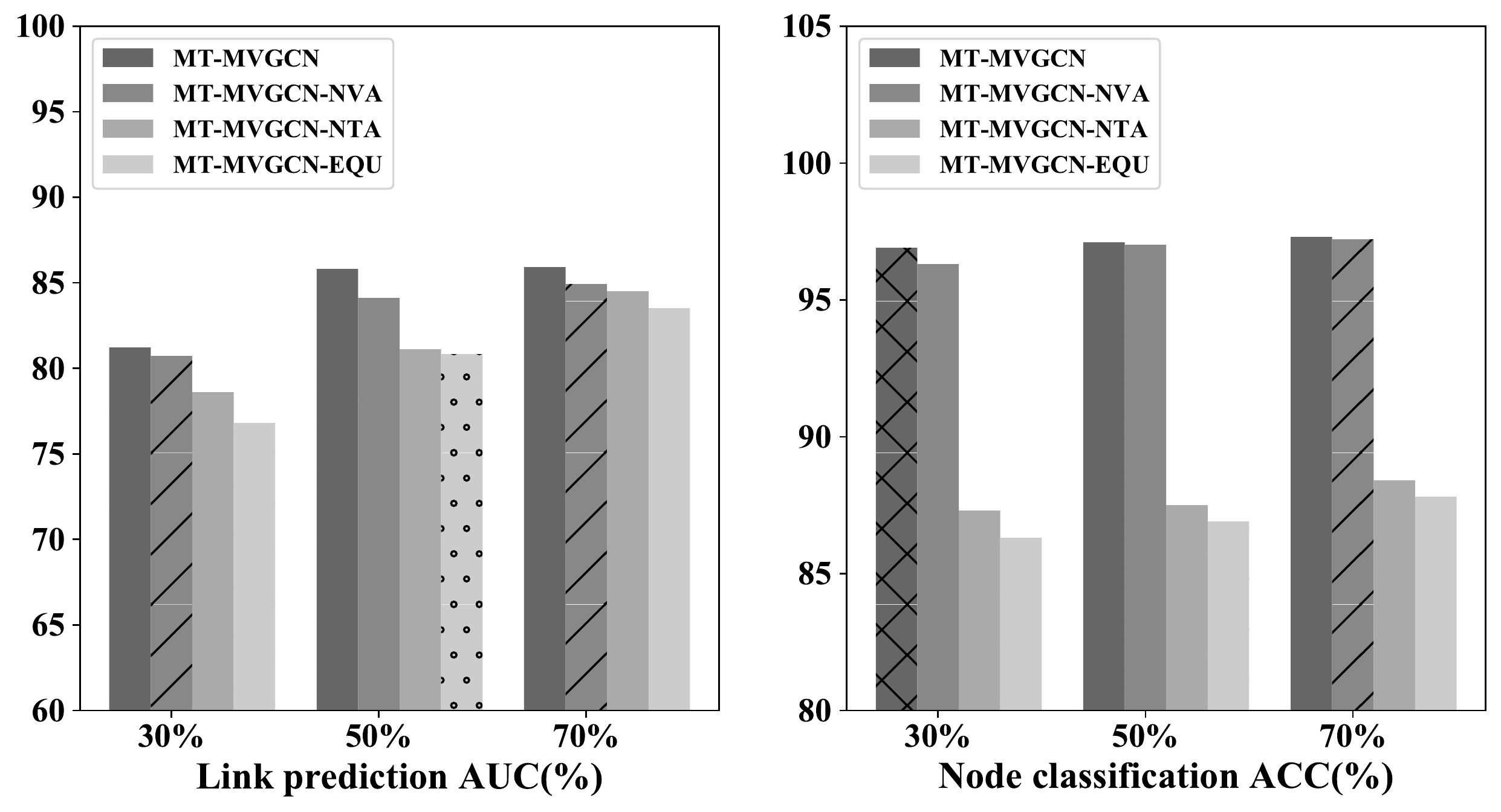}
    }
    \caption{Comparison of performance with different attention mechanisms. MT-MVGCN-NVA: MT-MVGCN without view attention mechanism, MT-MVGCN-NTA: MT-MVGCN without task attention mechanism, MT-MVGCN-EQU: MT-MVGCN with equal weights for views.}
    \label{fig:case_att}
\end{figure}

\subsection{Ablation Study of Attention Mechanisms}\label{att_case}

There are two important attention mechanisms in the proposed methods, the view attention mechanism and task attention mechanism. In this subsection, we evaluate the contributions of these two attention mechanisms to the final results. For each dataset, we first remove the view attention mechanism or task attention mechanism from MT-MVGCN model, namely as MT-MVGCN-NVA and MT-MVGCN-NTA respectively. Apart from that, we assign equal weights for views without attention mechanisms, named as MT-MVGCN-EQU. Then we study the link prediction and node classification results with different training ratios. And the results are presented in Figure ~\ref{fig:case_att}.

Overall, it is obvious that view attention mechanism and task attention mechanism are essential parts of the proposed model. For all datasets, there is a clear performance decline after we remove two attention mechanisms with the variation of the proportion of training data. This phenomenon shows that, rather than simply assigning equal weight to each view, the attention mechanisms make the model focus on the most informative views. Apart from that, link prediction task is more sensitive to the attention mechanisms than node classification task. As we can see, the performance of link prediction is marked retrogressive after the removing of attention mechanisms while node classification accuracy can still keep relative stable in Flickr dataset and only degrade on AMiner dataset.

And we still notice that the importance of task attention mechanism is beyond view attention mechanism. First, there is a larger performance decline under the condition of removing task attention mechanism. We can see that AUC and AP metrics are declined by 5\% on average if we remove task attention rather than view attention. Second, for link prediction task, the metrics of MT-MVGCN-NVA declines apparently on all datasets but still not sharply as MT-MVGCN-NTA. For node classification task, we even find that there are little change without view attention on Flickr dataset. By comparison, it is so clear for the declining in performance without task attention especially on AMiner dataset.

\begin{table}[tbp]
\centering
\setlength{\tabcolsep}{1.5mm}
\renewcommand\arraystretch{1.2}
\topcaption{Results of ablation studies of multi-task learning for the proposed model. We perform each task separately, namely 'Single' and perform all tasks via multi-task learning, namely 'multi'. Time cost is the average time cost per epoch and it is the sum of tasks for single-task learning.}
\label{single-multi}
\begin{tabular}{c|c|cc|cc}
\hline
\multicolumn{2}{c|}{\textbf{Dataset}}                  & \multicolumn{2}{c|}{\textbf{Flickr}} & \multicolumn{2}{c}{\textbf{AMiner}} \\ \hline
\textbf{Task}                        & \textbf{Metric} & \textbf{Single}   & \textbf{Multi}   & \textbf{Single}   & \textbf{Multi}   \\ \hline
\multirow{2}{*}{Link prediction}     & AP              & 0.820             & \textbf{0.847}   & 0.851             & \textbf{0.855}   \\
                                     & AUC             & 0.857             & \textbf{0.955}   & 0.892             & \textbf{0.897}   \\ \hline
\multirow{3}{*}{Node classification} & ACC             & \textbf{0.925}    & 0.922            & 0.975             & \textbf{0.977}   \\
                                     & Precision       & 0.789             & \textbf{0.800}   & 0.963             & \textbf{0.970}   \\
                                     & F1              & 0.772             & \textbf{0.780}   & 0.958             & \textbf{0.965}   \\ \hline
Time cost                            & Second          & 212               & \textbf{111}     & 0.369             & \textbf{0.284}   \\ \hline
\end{tabular}
\end{table}

\subsection{Ablation Study of Multi-task Learning}

Although our experimental results have demonstrated the effectiveness of multi-task learning, we here study the difference between training link prediction and node classification task separately and training them together for the proposed model on Flickr and AMiner datasets. We set the train data percentage as $0.5$ and keep other experimental settings the same including hardware\footnote{A shared device with Intel(R) Xeon(R) CPU E5-2680, a Tesla P100 GPU and 250Gb Memory.}.  The results are shown in Table ~\ref{single-multi}.

As we can see, the multi-task learning based results still outperform the single-task based results on two datasets. For link prediction task, the multi-task learning achieves better performance by 3\% on average, which means the information of node labels may be helpful for link prediction task. Besides, we still find that the influence varies for different datasets but it always positive. For node classification, we conclude that multi-task learning only brings a little influence due to the tiny difference between training tasks separately and simultaneously. Since it will not lead to the performance degradation, therefore training node classification with other tasks can be time efficient. As the graph convolutional networks tend to over-fitting resulting in unsatisfactory performance, we deduce that through multi-task learning the model is forced to learn more robust feature representations so the results are better on the whole. First, the model must learn general features for different tasks so the possibility of over-fitting is reduced. Second, different tasks can not only provide some useful information to each other, but also introduce some noise to each other to prevent over-fitting.

There is a key point that multi-task learning is very time efficient especially for the large networks. Flickr network is several times the size of AMiner network, thus we can observe that multi-task learning saves nearly double the time cost on Flickr dataset but for the AMiner dataset it is not so obvious. As a result, we can conclude that multi-task learning is time efficient and avoids a lot of repetitive work compared with single-task learning.

\subsection{Visualization of Attention Mechanisms}
\begin{figure*}[t]
    \centering
    \subfigure[\textbf{YouTube}]{
    \includegraphics[scale=0.42]{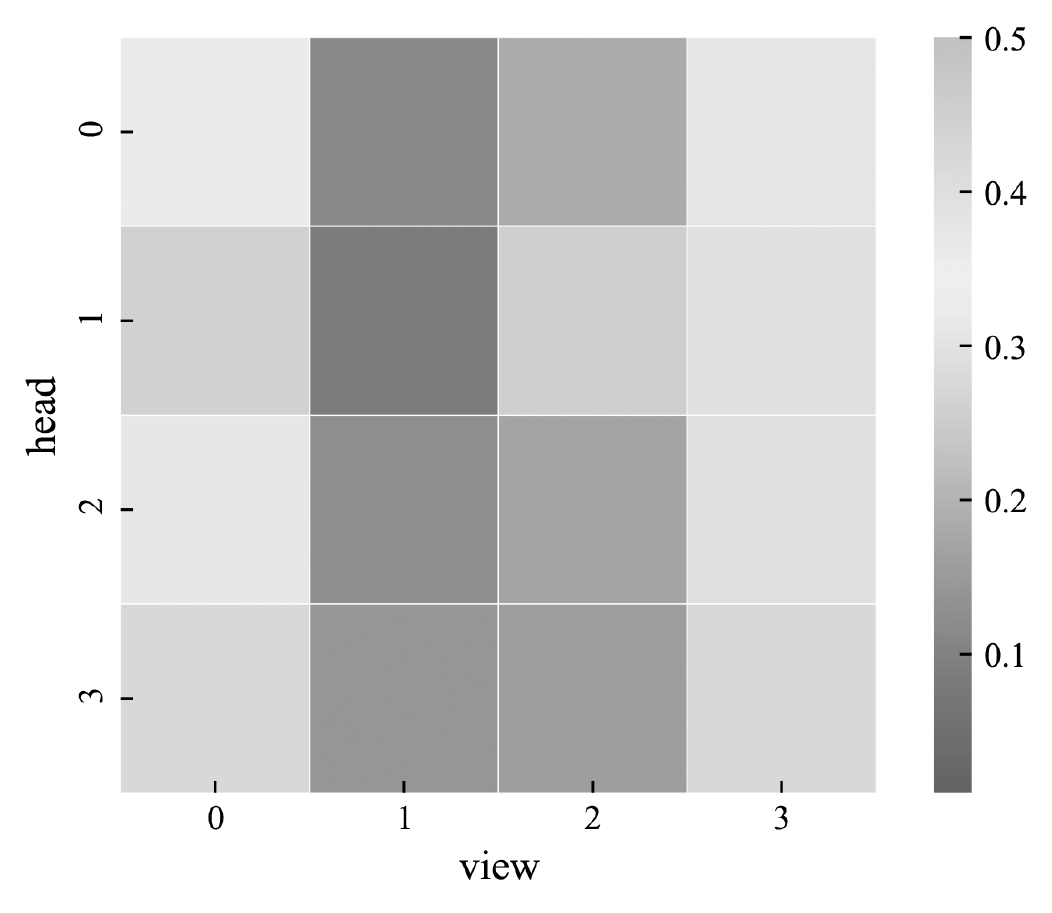}
    }
    \subfigure[\textbf{Twitter}]{
    \includegraphics[scale=0.42]{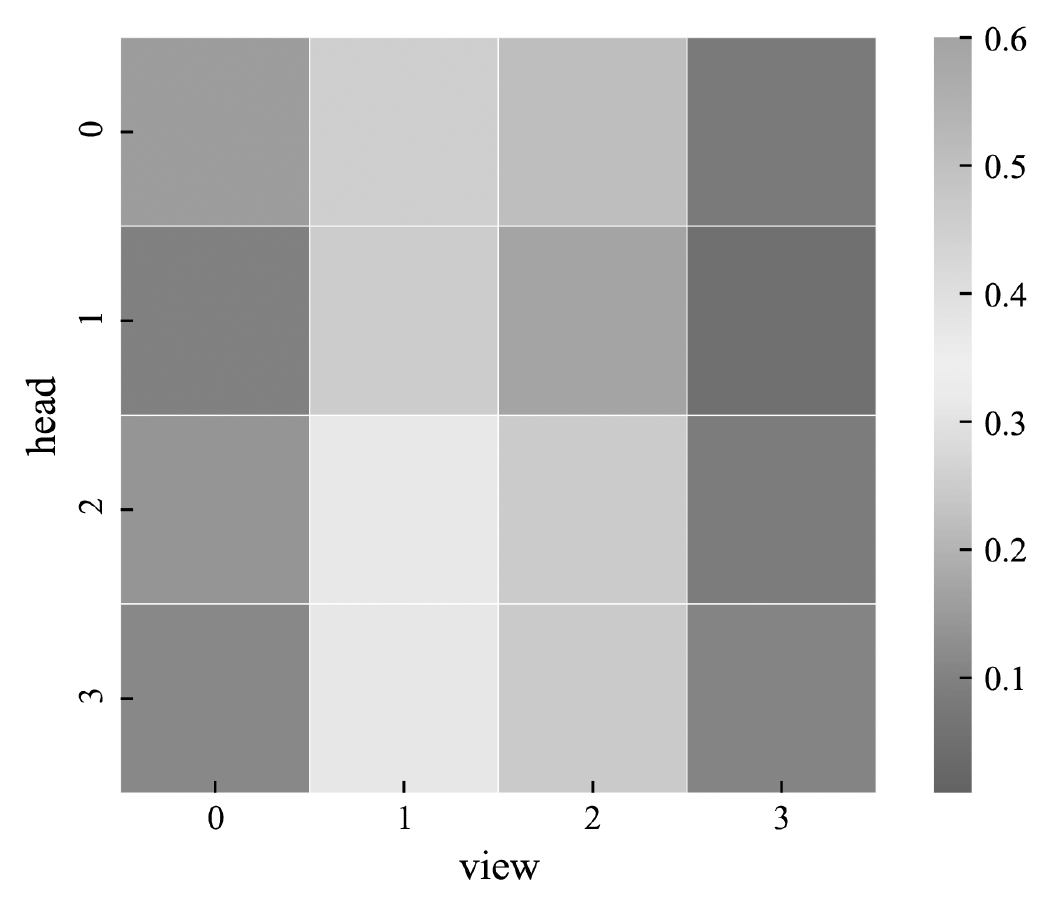}
    }
    \subfigure[\textbf{Flickr}]
    {
    \includegraphics[scale=0.42]{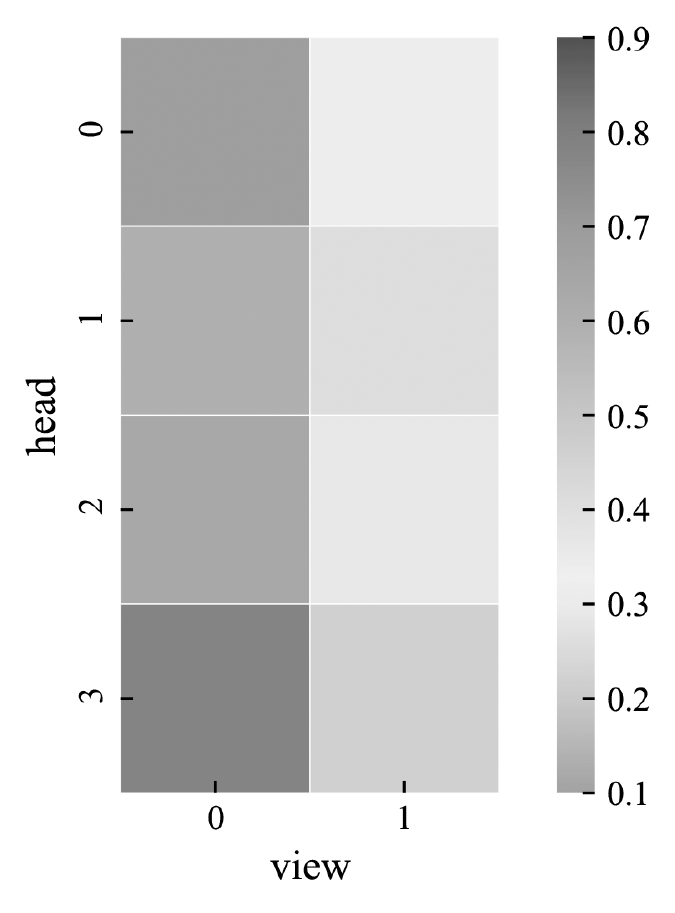}
    }
    \subfigure[\textbf{AMiner}]
    {
    \includegraphics[scale=0.42]{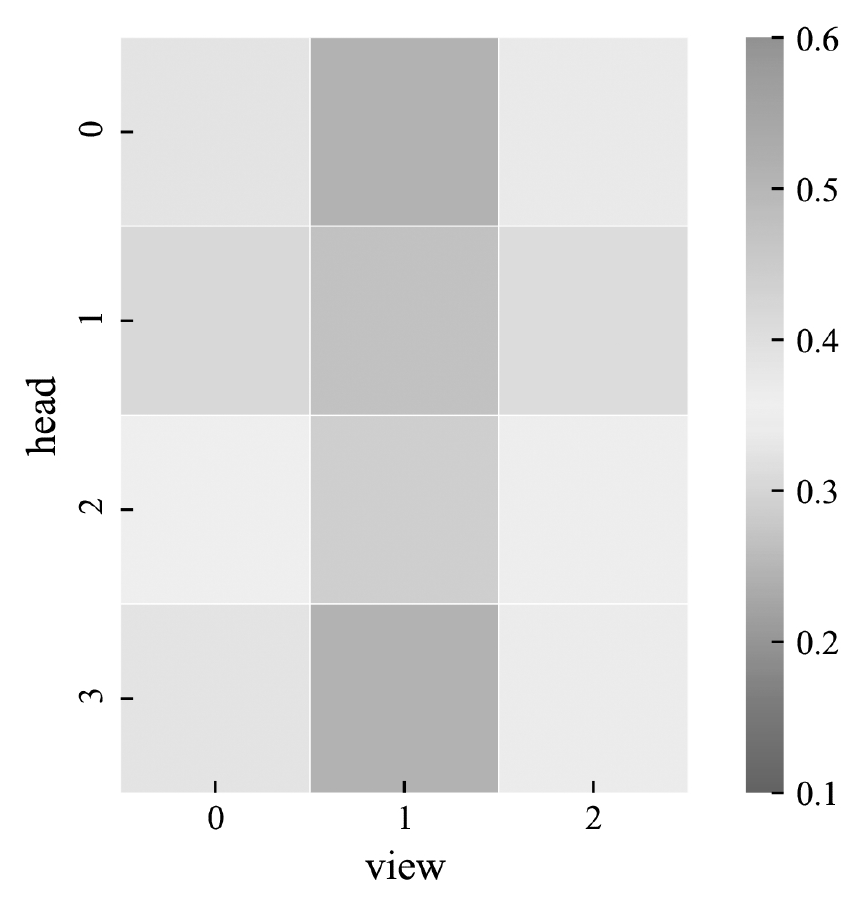}
    }
    \caption{Visualization of view weights learned by view attention mechanism of each dataset}
    \label{vis_v_att}
\end{figure*}

\begin{figure*}[t]
    \centering
    \subfigure[\textbf{YouTube}]{
    \includegraphics[scale=0.42]{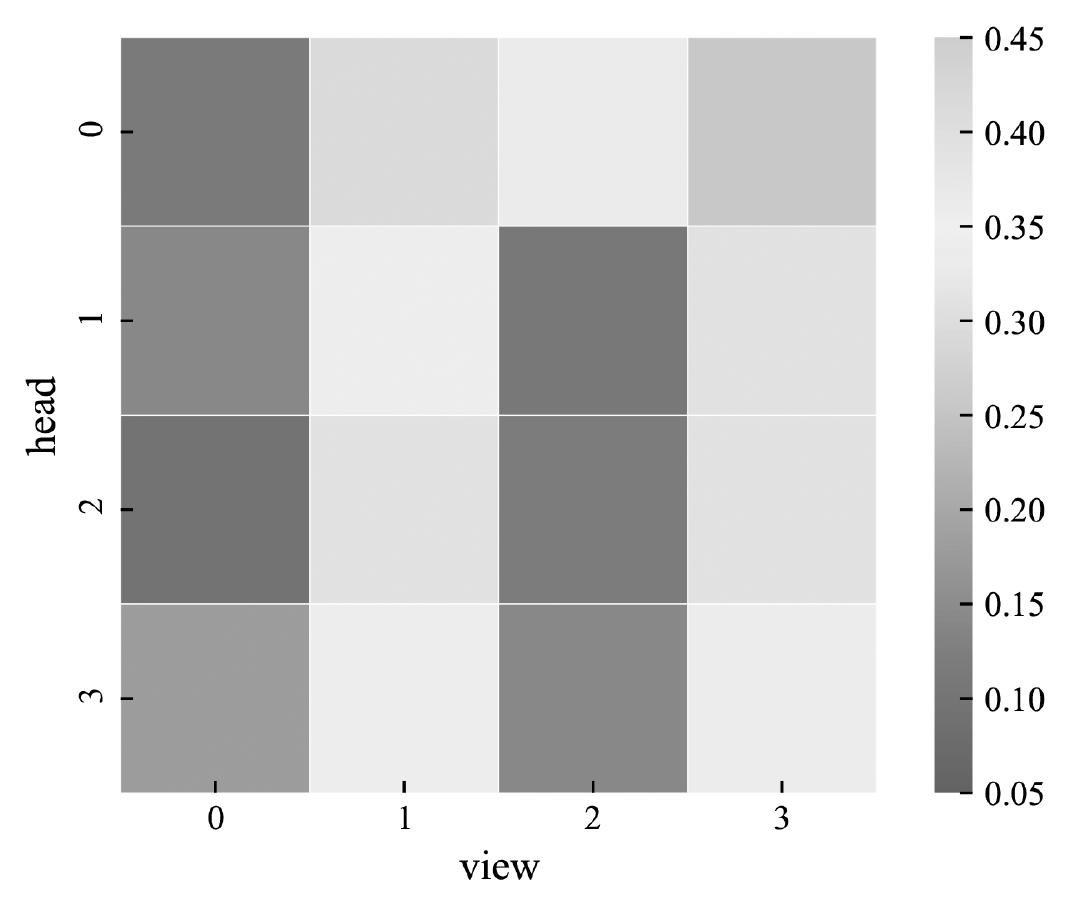}
    }
    \subfigure[\textbf{Twitter}]{
    \includegraphics[scale=0.42]{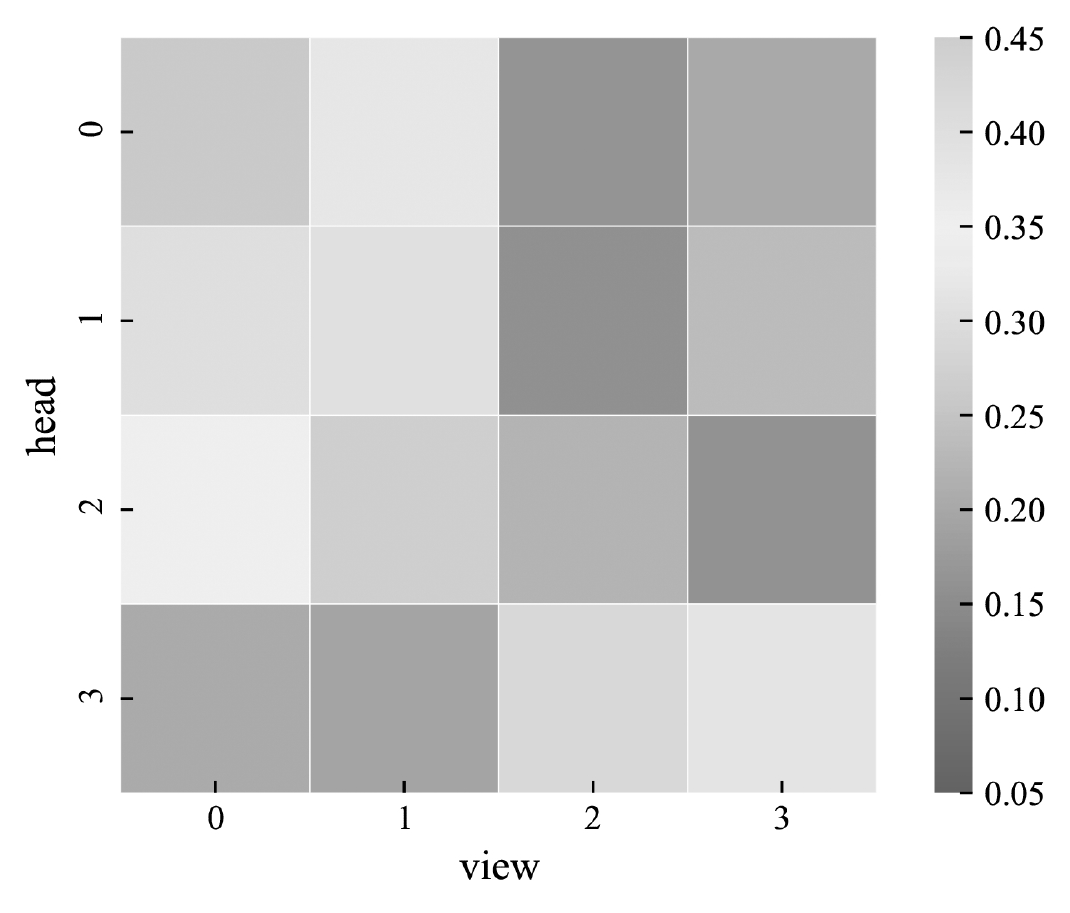}
    }
    \subfigure[\textbf{Flickr}]
    {
    \includegraphics[scale=0.42]{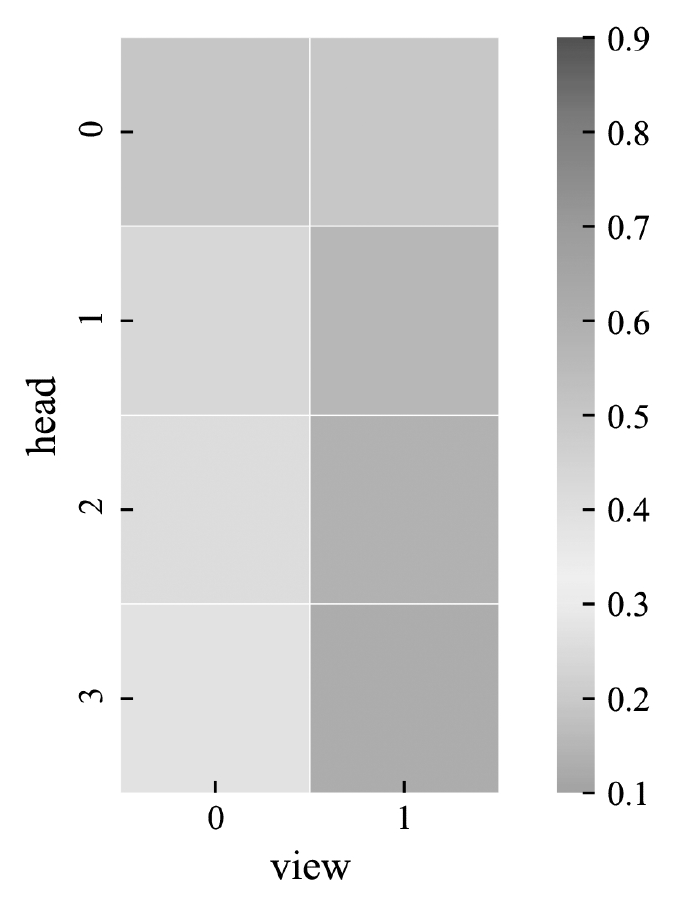}
    }
    \subfigure[\textbf{AMiner}]
    {
    \includegraphics[scale=0.42]{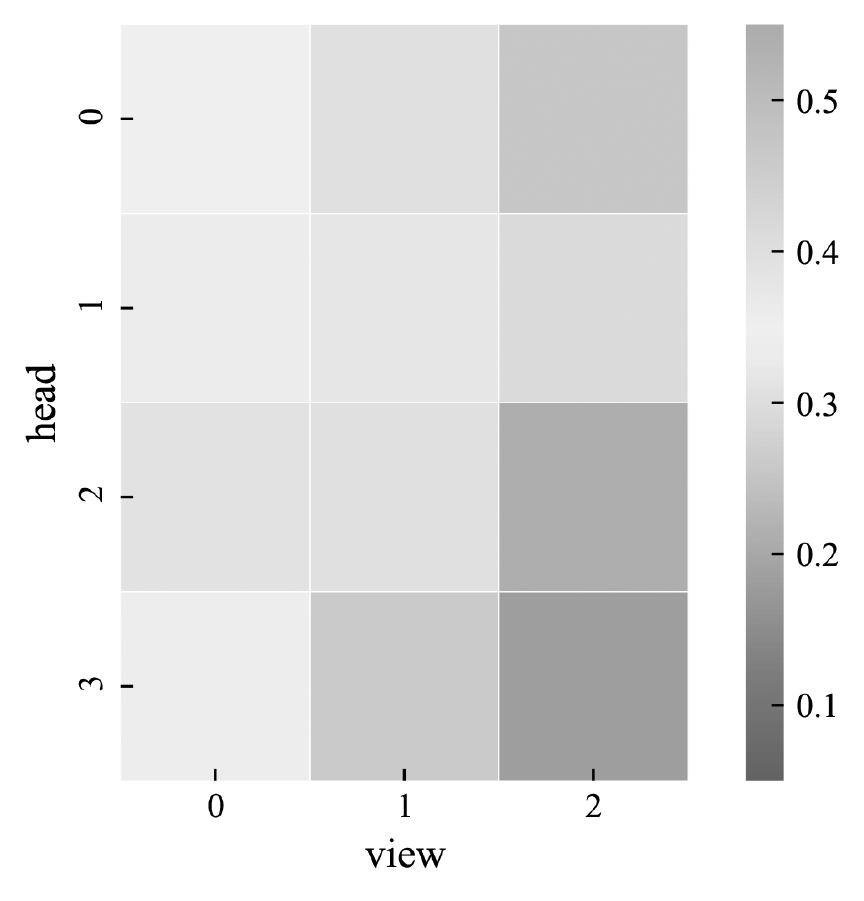}
    }
    \caption{Visualization of view weights learned by link prediction task attention mechanism of each dataset}
    \label{vis_l_att}
\end{figure*}

\begin{figure}[t]
    \centering
    \subfigure[\textbf{Flickr}]
    {
    \includegraphics[scale=0.42]{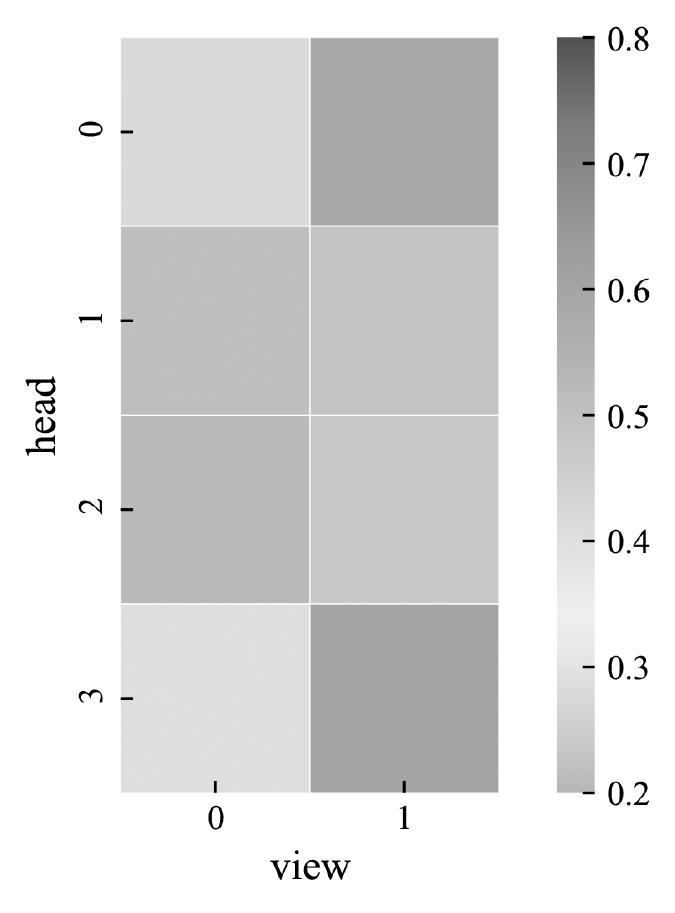}
    }
    \subfigure[\textbf{AMiner}]
    {
    \includegraphics[scale=0.42]{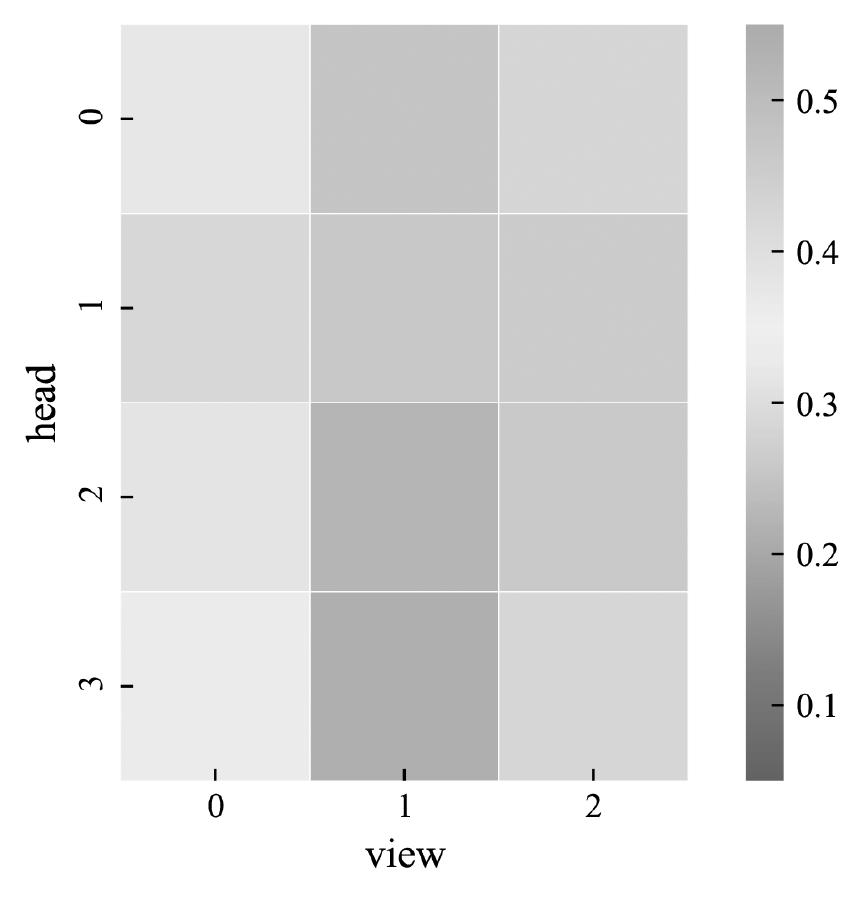}
    }
    \caption{Visualization of view weights learned by node classification task attention mechanism of AMiner dataset and Flickr dataset}
    \label{vis_n_att}
\end{figure}

To validate the effectiveness of the view and task attention mechanisms, we explore the learned weight for each view of different datasets. Now that we have touched upon attention heads, we also show where the different heads are focusing. The visualizations of view attention weights, link prediction task attention weights and node classification task attention weights are shown in Figure~\ref{vis_v_att}, \ref{vis_l_att} and \ref{vis_n_att} respectively. As we can see, the weights of multiple views have noticeable difference between each other, which means that the view attention mechanism try to assign proper importance to each view. For example, the third view of YouTube dataset is more valued and the second view of Twitter view has relative large weight than other views. Moreover, we also observe that dividing the attention mechanism into multiple heads and forming multiple sub-spaces allows it to focus on different aspects of information. Compared with view attention mechanism, the task attention mechanism assigns different weight to each view since different tasks tend to utilize distinct view information to achieve better performance. Intuitively, view attention mechanism are shared by all tasks so it would like to extract consensus information while task attention mechanism prefers to extract task-specific information, leading to different weights learned by this two kinds of attention mechanisms. Overall, our model can achieve better performance via two kinds attention mechanisms capturing consensus information as well as task-specific information.

\begin{figure*}[t]
    \centering
    \subfigure[\textbf{Flickr}]
    {
    \includegraphics[scale=0.42]{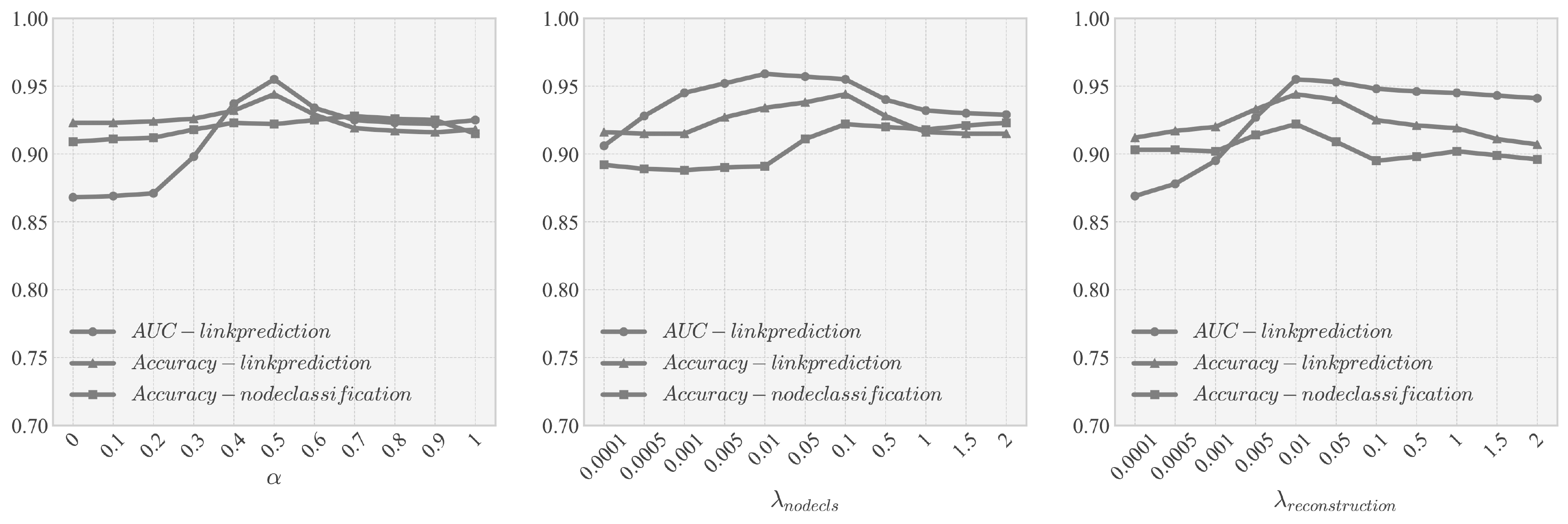}
    }
    \\
    \subfigure[\textbf{AMiner}]
    {
    \includegraphics[scale=0.42]{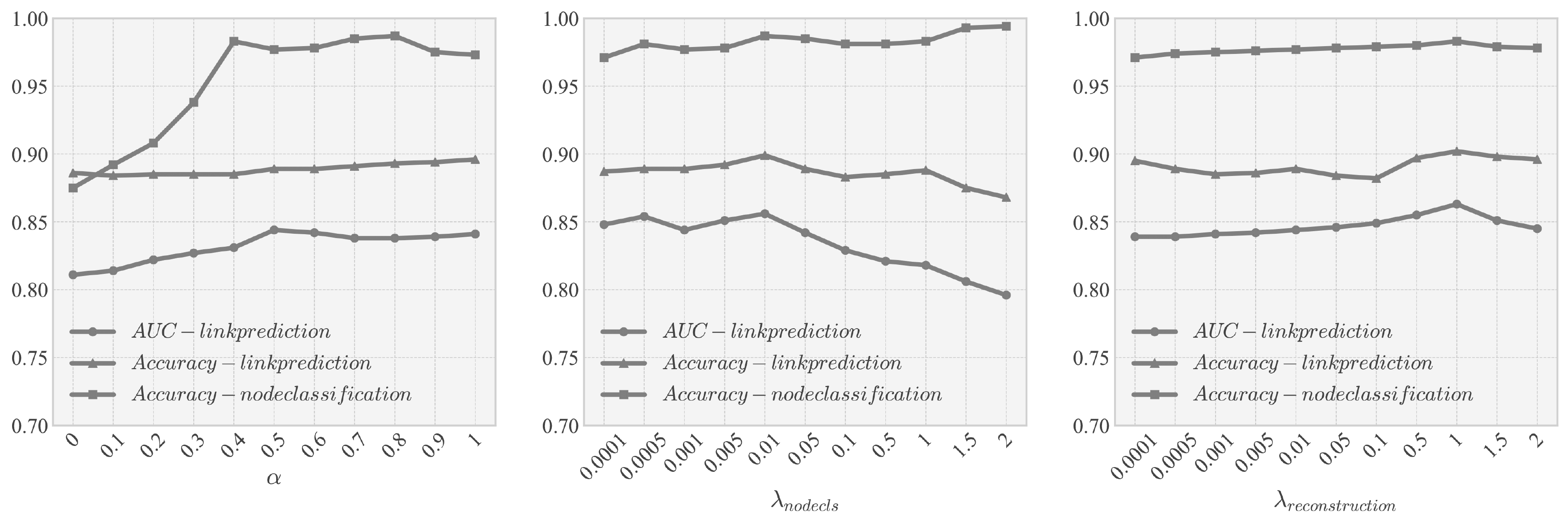}
    }
    \caption{Parameter Sensitivity w.r.t. the balance of two attention mechanisms $\alpha$, the importance of node classification task $\lambda_{nodecls}$ and the importance of view reconstruction task $\lambda_{reconstruction}$.}
    \label{parameter}
\end{figure*}

\subsection{Parameter Sensitivity}
In this section, we study the parameter sensitivity of the proposed model. There are three important parameters, the balance between two attention mechanisms $\alpha$, the importance of node classification task $\lambda_{nodecls}$, and the importance of view reconstruction task $\lambda_{reconstruction}$. As we keep the importance of link prediction task as $1.0$, we only need to vary the importance of other two tasks. The training percentage is set as $0.5$. We choose Flickr and AMiner datasets as examples and the results are shown in Figure ~\ref{parameter}.

\subsubsection{\textbf{The influence of $\alpha$}} We first analyze how the balance of two attention mechanisms affects the experimental results. For both tasks we can see that the performance initially raises with the $\alpha$ increasing. The link prediction accuracy raises a little while the AUC raises sharply on both datasets. For node classification task, the performance on AMiner dataset is more sensitive to the attention mechanisms than its on Flickr dataset. Then with the increasing of $\alpha$, the performance of both tasks keeps relatively stable without decline for two tasks. In addition, corresponding to section~\ref{att_case}, the task attention mechanism is key factor of the proposed models. If the weight of task attention mechanism is too low the performance is also poor. On the whole, we conclude that keeping the importance of two attention mechanisms roughly equal is a favourable setting for the model.

\subsubsection{\textbf{The influence of $\lambda_{nodecls}$}} Then we discuss the influence of node classification task, with the importance of it varying from $0.0001$ to $2.0$. As observed in Figure~\ref{parameter}, there is little difference in the effect on the two datasets. For Flickr dataset, when the hyper-parameter is too small or too large, the link prediction results are not very satisfactory. This phenomenon may due to the reason that if the importance of node classification task is too low the link prediction task can not obtain enough information; while the importance is too high the model will ignores the link prediction task, thus it is not surprised that the performance decline. Compared with Flickr dataset, there is no obvious change in link prediction performance when $\lambda_{nodecls}$ is small on AMiner dataset. However, Flickr dataset is faced with the performance degradation in link prediction task if the parameter is not properly set. For node classification task, the influence is basically the same as link prediction task. With the increasing of importance the model start focus more on node classification gradually thus the classification accuracy continues growing.

\subsubsection{\textbf{The influence of $\lambda_{reconstruction}$}} Finally we study the impact of $\lambda_{reconstruction}$, which controls the contribution of reconstruction task to our model. As we can see, the auxiliary task is pretty significant for our model especially for Flickr dataset. The performance for both tasks raises evidently with the weight of view reconstruction increasing at start. Then there is a slightly degradation of performance if the weight of view reconstruction sustained increases. In contrast, it seems that link prediction task is not sensitive to the auxiliary task on AMiner dataset, but it has no negative impact on performance as long as the weight is set appropriately. All in all, our model boosts the performance in most cases through the view reconstruction as it helps to preserve structural information of different views.

\subsection{Discussion}

In this subsection, we mainly discuss the spotlights of our models and try to explain why our models perform better than experimental baselines. 

Compared with existing multi-view based methods, such as average, concatenate and add, our models achieve great results due to the advanced view fusion strategy. First, we propose an attention-based fusion method which can capture the complex relationships between views. We not only allow all views to vote for the robust representation, but also allow the specific task to focus on the most informative views. Second, as mentioned in~\cite{xu2018powerful}, average fails to capture repeat features in graph, thus average methods and something like that may also fail to fuse different views information since it is common for repeating features among views. Third, add and concatenate methods are enough to preserve the original information, but may introduce extra noise and lack sufficient collaboration of views, making the results sub-optimal. Finally, previous attention based methods like MVE~\cite{meng2017ention} adopts simple attention mechanism to take an average of all views, which is not interpretable and effective as our models. In contrast, by applying two multi-head attention mechanisms our models can also compute diverse attentions independently on different sub-spaces in parallel.

Compared with the existing multi-task based method, i.e. MTGAE, our models include shared layers and task specific layers, which brings great flexibility to multi-task learning. MTGAE~\cite{Tran-LoNGAE:2018} is composed of a symmetrical graph auto-encoder shared by all tasks so it is hard to extract specific information for a certain task. The advantage of our models lies in that view attention is shared by tasks and task attention is specific to each task, expected to capture more complex relationships between network information and tasks. In addition, the auxiliary task also makes great contributions to multi-task learning.

\section{Conclusion} \label{conclusion}
In this paper, we study the problem of link prediction and node classification tasks by multi-task learning with multi-view graph convolutional networks. More specifically, our model considers utilizing information of multiple views for multiple tasks through two attention mechanisms. Besides, we enhance our model by introducing an auxiliary task which can make the network structural information better preserved. In the experiments, we show that our model is significantly and consistently superior to the start-of-the-art baselines. We also study the contribution of two attention mechanisms and the parameter sensitivity. In the future, we plan to explore two directions based on the current work: one is how to make the model learn the importance of different tasks by itself rather than controlling it by hyper-parameters. The other is applying our model to more realistic scenarios and problems.

\section*{Acknowledgement}
Hong Huang is supported by National Natural Science Foundation of China (No. 61802140).

\bibliographystyle{IEEEtran}
\bibliography{sample-base}

\begin{thebibliography}{10}
\providecommand{\url}[1]{#1}
\csname url@samestyle\endcsname
\providecommand{\newblock}{\relax}
\providecommand{\bibinfo}[2]{#2}
\providecommand{\BIBentrySTDinterwordspacing}{\spaceskip=0pt\relax}
\providecommand{\BIBentryALTinterwordstretchfactor}{4}
\providecommand{\BIBentryALTinterwordspacing}{\spaceskip=\fontdimen2\font plus
\BIBentryALTinterwordstretchfactor\fontdimen3\font minus
  \fontdimen4\font\relax}
\providecommand{\BIBforeignlanguage}[2]{{%
\expandafter\ifx\csname l@#1\endcsname\relax
\typeout{** WARNING: IEEEtran.bst: No hyphenation pattern has been}%
\typeout{** loaded for the language `#1'. Using the pattern for}%
\typeout{** the default language instead.}%
\else
\language=\csname l@#1\endcsname
\fi
#2}}
\providecommand{\BIBdecl}{\relax}
\BIBdecl

\bibitem{tang2008arnetminer}
J.~Tang, J.~Zhang, L.~Yao, J.~Li, L.~Zhang, and Z.~Su, ``Arnetminer: extraction
  and mining of academic social networks,'' in \emph{Proceedings of the 14th
  ACM SIGKDD international conference on Knowledge discovery and data
  mining}.\hskip 1em plus 0.5em minus 0.4em\relax ACM, 2008, pp. 990--998.

\bibitem{li2017scored}
T.~Li, R.~Wernersson, R.~B. Hansen, H.~Horn, J.~Mercer, G.~Slodkowicz, C.~T.
  Workman, O.~Rigina, K.~Rapacki, H.~H. St{\ae}rfeldt \emph{et~al.}, ``A scored
  human protein--protein interaction network to catalyze genomic
  interpretation,'' \emph{Nature methods}, vol.~14, no.~1, p.~61, 2017.

\bibitem{van2014online}
R.~Van~Noorden, ``Online collaboration: Scientists and the social network,''
  \emph{Nature news}, vol. 512, no. 7513, p. 126, 2014.

\bibitem{romero2016social}
D.~M. Romero, B.~Uzzi, and J.~Kleinberg, ``Social networks under stress,'' in
  \emph{Proceedings of the 25th International Conference on World Wide
  Web}.\hskip 1em plus 0.5em minus 0.4em\relax International World Wide Web
  Conferences Steering Committee, 2016, pp. 9--20.

\bibitem{cui2018survey}
P.~Cui, X.~Wang, J.~Pei, and W.~Zhu, ``A survey on network embedding,''
  \emph{IEEE Transactions on Knowledge and Data Engineering}, vol.~31, no.~5,
  pp. 833--852, 2018.

\bibitem{perozzi2014deepwalk}
B.~Perozzi, R.~Al-Rfou, and S.~Skiena, ``Deepwalk: Online learning of social
  representations,'' in \emph{Proceedings of the 20th ACM SIGKDD international
  conference on Knowledge discovery and data mining}.\hskip 1em plus 0.5em
  minus 0.4em\relax ACM, 2014, pp. 701--710.

\bibitem{mikolov2013distributed}
T.~Mikolov, I.~Sutskever, K.~Chen, G.~S. Corrado, and J.~Dean, ``Distributed
  representations of words and phrases and their compositionality,'' in
  \emph{Proceedings of the 27th Conference on Advances in neural information
  processing systems}, 2013, pp. 3111--3119.

\bibitem{tang2015line}
J.~Tang, M.~Qu, M.~Wang, M.~Zhang, J.~Yan, and Q.~Mei, ``Line: Large-scale
  information network embedding,'' in \emph{Proceedings of World Wide Web
  Conference}, 2015, pp. 1067--1077.

\bibitem{wang2016structural}
D.~Wang, P.~Cui, and W.~Zhu, ``Structural deep network embedding,'' in
  \emph{Proceedings of the 22nd ACM SIGKDD international conference on
  Knowledge discovery and data mining}.\hskip 1em plus 0.5em minus 0.4em\relax
  ACM, 2016, pp. 1225--1234.

\bibitem{ribeiro2017struc2vec}
L.~F. Ribeiro, P.~H. Saverese, and D.~R. Figueiredo, ``struc2vec: Learning node
  representations from structural identity,'' in \emph{Proceedings of the 23rd
  ACM SIGKDD International Conference on Knowledge Discovery and Data
  Mining}.\hskip 1em plus 0.5em minus 0.4em\relax ACM, 2017, pp. 385--394.

\bibitem{grover2016node2vec}
A.~Grover and J.~Leskovec, ``node2vec: Scalable feature learning for
  networks,'' in \emph{Proceedings of the 22nd ACM SIGKDD international
  conference on Knowledge discovery and data mining}.\hskip 1em plus 0.5em
  minus 0.4em\relax ACM, 2016, pp. 855--864.

\bibitem{defferrard2016convolutional}
M.~Defferrard, X.~Bresson, and P.~Vandergheynst, ``Convolutional neural
  networks on graphs with fast localized spectral filtering,'' in
  \emph{Proceedings of the 30th Conference on Advances in neural information
  processing systems}, 2016, pp. 3844--3852.

\bibitem{bruna2013spectral}
J.~Bruna, W.~Zaremba, A.~Szlam, and Y.~LeCun, ``Spectral networks and locally
  connected networks on graphs,'' in \emph{Proceedings of the 2nd International
  Conference on Learning Representations, {ICLR} 2014, Banff, AB, Canada, April
  14-16, 2014, Conference Track Proceedings}, Y.~Bengio and Y.~LeCun, Eds.,
  2014.

\bibitem{kipf2016semi}
T.~N. Kipf and M.~Welling, ``Semi-supervised classification with graph
  convolutional networks,'' in \emph{Proceedings of the 5th International
  Conference on Learning Representations, {ICLR} 2017, Toulon, France, April
  24-26, 2017, Conference Track Proceedings}.\hskip 1em plus 0.5em minus
  0.4em\relax OpenReview.net, 2017.

\bibitem{meng2017ention}
J.~T. Meng, J.~Shang, X.~Ren, M.~Zhang, and J.~Han, ``An attention-based
  collaboration framework for multi-view network representation learning,'' in
  \emph{Proceedings of the 2017 {ACM} on Conference on Information and
  Knowledge Management, {CIKM} 2017, Singapore, November 06 - 10, 2017}.\hskip
  1em plus 0.5em minus 0.4em\relax {ACM}, 2017, pp. 1767--1776.

\bibitem{ma2018multi}
Y.~Ma, Z.~Ren, Z.~Jiang, J.~Tang, and D.~Yin, ``Multi-dimensional network
  embedding with hierarchical structure,'' in \emph{Proceedings of the Eleventh
  ACM International Conference on Web Search and Data Mining}.\hskip 1em plus
  0.5em minus 0.4em\relax ACM, 2018, pp. 387--395.

\bibitem{zhao2017multi}
H.~Zhao, Z.~Ding, and Y.~Fu, ``Multi-view clustering via deep matrix
  factorization.'' in \emph{Proceedings of the Thirty-First {AAAI} Conference
  on Artificial Intelligence, February 4-9, 2017, San Francisco, California,
  {USA}}, 2017, pp. 2921--2927.

\bibitem{sun2013survey}
S.~Sun, ``A survey of multi-view machine learning,'' \emph{Neural Computing and
  Applications}, vol.~23, no. 7-8, pp. 2031--2038, 2013.

\bibitem{shi2018mvn2vec}
Y.~Shi, F.~Han, X.~He, X.~He, C.~Yang, J.~Luo, and J.~Han, ``mvn2vec:
  Preservation and collaboration in multi-view network embedding,'' \emph{arXiv
  preprint arXiv:1801.06597}, 2018.

\bibitem{niwattanakul2013using}
S.~Niwattanakul, J.~Singthongchai, E.~Naenudorn, and S.~Wanapu, ``Using of
  jaccard coefficient for keywords similarity,'' in \emph{Proceedings of the
  international multiconference of engineers and computer scientists}, vol.~1,
  no.~6, 2013, pp. 380--384.

\bibitem{wang2017community}
X.~Wang, P.~Cui, J.~Wang, J.~Pei, W.~Zhu, and S.~Yang, ``Community preserving
  network embedding,'' in \emph{Proceedings of the Thirty-First AAAI Conference
  on Artificial Intelligence}.\hskip 1em plus 0.5em minus 0.4em\relax {AAAI}
  Press, 2017, pp. 203--209.

\bibitem{qiu2018network}
J.~Qiu, Y.~Dong, H.~Ma, J.~Li, K.~Wang, and J.~Tang, ``Network embedding as
  matrix factorization: Unifying deepwalk, line, pte, and node2vec,'' in
  \emph{Proceedings of the Eleventh ACM International Conference on Web Search
  and Data Mining}.\hskip 1em plus 0.5em minus 0.4em\relax ACM, 2018, pp.
  459--467.

\bibitem{lee2001algorithms}
D.~D. Lee and H.~S. Seung, ``Algorithms for non-negative matrix
  factorization,'' in \emph{Proceedings of the 15th Conference on Advances in
  neural information processing systems}, 2001, pp. 556--562.

\bibitem{cao2016deep}
S.~Cao, W.~Lu, and Q.~Xu, ``Deep neural networks for learning graph
  representations,'' in \emph{Proceedings of the Thirtieth AAAI Conference on
  Artificial Intelligence}.\hskip 1em plus 0.5em minus 0.4em\relax {AAAI}
  Press, 2016, pp. 1145--1152.

\bibitem{dong2017metapath2vec}
Y.~Dong, N.~V. Chawla, and A.~Swami, ``metapath2vec: Scalable representation
  learning for heterogeneous networks,'' in \emph{Proceedings of the 23rd ACM
  SIGKDD international conference on knowledge discovery and data
  mining}.\hskip 1em plus 0.5em minus 0.4em\relax ACM, 2017, pp. 135--144.

\bibitem{fu2017hin2vec}
T.-y. Fu, W.-C. Lee, and Z.~Lei, ``Hin2vec: Explore meta-paths in heterogeneous
  information networks for representation learning,'' in \emph{Proceedings of
  the 2017 ACM on Conference on Information and Knowledge Management}.\hskip
  1em plus 0.5em minus 0.4em\relax ACM, 2017, pp. 1797--1806.

\bibitem{wang2018shine}
H.~Wang, F.~Zhang, M.~Hou, X.~Xie, M.~Guo, and Q.~Liu, ``Shine: Signed
  heterogeneous information network embedding for sentiment link prediction,''
  in \emph{Proceedings of the Eleventh ACM International Conference on Web
  Search and Data Mining}.\hskip 1em plus 0.5em minus 0.4em\relax ACM, 2018,
  pp. 592--600.

\bibitem{monti2017geometric}
F.~Monti, M.~Bronstein, and X.~Bresson, ``Geometric matrix completion with
  recurrent multi-graph neural networks,'' in \emph{Proceedings of the 31st
  Conference on Advances in Neural Information Processing Systems}, 2017, pp.
  3697--3707.

\bibitem{gilmer2017neural}
J.~Gilmer, S.~S. Schoenholz, P.~F. Riley, O.~Vinyals, and G.~E. Dahl, ``Neural
  message passing for quantum chemistry,'' in \emph{Proceedings of the 34th
  International Conference on Machine Learning-Volume 70}.\hskip 1em plus 0.5em
  minus 0.4em\relax JMLR. org, 2017, pp. 1263--1272.

\bibitem{velivckovic2017graph}
P.~Velickovic, G.~Cucurull, A.~Casanova, A.~Romero, P.~Li{\`{o}}, and
  Y.~Bengio, ``Graph attention networks,'' in \emph{Proceedings of the 6th
  International Conference on Learning Representations, {ICLR} 2018, Vancouver,
  BC, Canada, April 30 - May 3, 2018, Conference Track Proceedings}.\hskip 1em
  plus 0.5em minus 0.4em\relax OpenReview.net, 2018.

\bibitem{chen2018fastgcn}
J.~Chen, T.~Ma, and C.~Xiao, ``Fastgcn: fast learning with graph convolutional
  networks via importance sampling,'' in \emph{Proceedings of the 6th
  International Conference on Learning Representations, {ICLR} 2018, Vancouver,
  BC, Canada, April 30 - May 3, 2018, Conference Track Proceedings}, 2018.

\bibitem{wang2017signed}
S.~Wang, J.~Tang, C.~Aggarwal, Y.~Chang, and H.~Liu, ``Signed network embedding
  in social media,'' in \emph{Proceedings of the 2017 SIAM international
  conference on data mining}.\hskip 1em plus 0.5em minus 0.4em\relax SIAM,
  2017, pp. 327--335.

\bibitem{li2017deepcas}
C.~Li, J.~Ma, X.~Guo, and Q.~Mei, ``Deepcas: An end-to-end predictor of
  information cascades,'' in \emph{Proceedings of the 26th international
  conference on World Wide Web}.\hskip 1em plus 0.5em minus 0.4em\relax
  International World Wide Web Conferences Steering Committee, 2017, pp.
  577--586.

\bibitem{tu2016max}
C.~Tu, W.~Zhang, Z.~Liu, M.~Sun \emph{et~al.}, ``Max-margin deepwalk:
  Discriminative learning of network representation.'' in \emph{Proceedings of
  the Twenty-Fifth International Joint Conference on Artificial Intelligence,
  {IJCAI} 2016, New York, NY, USA, 9-15 July 2016}, 2016, pp. 3889--3895.

\bibitem{zhang2014facial}
Z.~Zhang, P.~Luo, C.~C. Loy, and X.~Tang, ``Facial landmark detection by deep
  multi-task learning,'' in \emph{Proceedings of the 13th European conference
  on computer vision}.\hskip 1em plus 0.5em minus 0.4em\relax Springer, 2014,
  pp. 94--108.

\bibitem{trivedi2010multiview}
A.~Trivedi, P.~Rai, H.~Daum{\'e}~III, and S.~L. DuVall, ``Multiview clustering
  with incomplete views,'' in \emph{NIPS Workshop}, vol. 224, 2010.

\bibitem{liu2017adversarial}
P.~Liu, X.~Qiu, and X.~Huang, ``Adversarial multi-task learning for text
  classification,'' in \emph{Proceedings of the 55th Annual Meeting of the
  Association for Computational Linguistics, {ACL} 2017, Vancouver, Canada,
  July 30 - August 4, Volume 1: Long Papers}, R.~Barzilay and M.~Kan,
  Eds.\hskip 1em plus 0.5em minus 0.4em\relax Association for Computational
  Linguistics, 2017, pp. 1--10.

\bibitem{Tran-LoNGAE:2018}
P.~V. Tran, ``Multi-task graph autoencoders,'' in \emph{Workshop on Relational
  Representation Learning, NIPS 2018, Montréal, Canada.}, 2018.

\bibitem{lu2017multilinear}
C.-T. Lu, L.~He, W.~Shao, B.~Cao, and P.~S. Yu, ``Multilinear factorization
  machines for multi-task multi-view learning,'' in \emph{Proceedings of the
  Tenth ACM International Conference on Web Search and Data Mining}.\hskip 1em
  plus 0.5em minus 0.4em\relax ACM, 2017, pp. 701--709.

\bibitem{zhang2016multi}
X.~Zhang, X.~Zhang, H.~Liu, and X.~Liu, ``Multi-task multi-view clustering,''
  \emph{IEEE Transactions on Knowledge and Data Engineering}, vol.~28, no.~12,
  pp. 3324--3338, 2016.

\bibitem{vaswani2017attention}
A.~Vaswani, N.~Shazeer, N.~Parmar, J.~Uszkoreit, L.~Jones, A.~N. Gomez,
  {\L}.~Kaiser, and I.~Polosukhin, ``Attention is all you need,'' in
  \emph{Proceedings of the 31th Conference on Advances in neural information
  processing systems}, 2017, pp. 5998--6008.

\bibitem{khattar2018hram}
D.~Khattar, V.~Kumar, V.~Varma, and M.~Gupta, ``Hram: A hybrid recurrent
  attention machine for news recommendation,'' in \emph{Proceedings of the 27th
  ACM International Conference on Information and Knowledge Management}.\hskip
  1em plus 0.5em minus 0.4em\relax ACM, 2018, pp. 1619--1622.

\bibitem{mnih2014recurrent}
V.~Mnih, N.~Heess, A.~Graves \emph{et~al.}, ``Recurrent models of visual
  attention,'' in \emph{Proceedings of the Twenty-eighth Conference on Advances
  in neural information processing systems}, 2014, pp. 2204--2212.

\bibitem{li2018deeper}
Q.~Li, Z.~Han, and X.-M. Wu, ``Deeper insights into graph convolutional
  networks for semi-supervised learning,'' in \emph{Proceedings of the
  Thirty-Second AAAI Conference}, 2018.

\bibitem{salakhutdinov2009semantic}
R.~Salakhutdinov and G.~Hinton, ``Semantic hashing,'' \emph{International
  Journal of Approximate Reasoning}, vol.~50, no.~7, pp. 969--978, 2009.

\bibitem{abadi2016tensorflow}
M.~Abadi, P.~Barham, J.~Chen, Z.~Chen, A.~Davis, J.~Dean, M.~Devin,
  S.~Ghemawat, G.~Irving, M.~Isard \emph{et~al.}, ``Tensorflow: A system for
  large-scale machine learning,'' in \emph{12th {USENIX} Symposium on Operating
  Systems Design and Implementation, {OSDI} 2016, Savannah, GA, USA, November
  2-4, 2016}, 2016, pp. 265--283.

\bibitem{zafarani2009social}
R.~Zafarani and H.~Liu, ``Social computing data repository at asu
  [http://socialcomputing. asu. edu]. tempe, az: Arizona state university,
  school of computing,'' \emph{Informatics and Decision Systems Engineering},
  2009.

\bibitem{de2013anatomy}
M.~De~Domenico, A.~Lima, P.~Mougel, and M.~Musolesi, ``The anatomy of a
  scientific rumor,'' \emph{Scientific reports}, vol.~3, p. 2980, 2013.

\bibitem{tang2009relational}
L.~Tang and H.~Liu, ``Relational learning via latent social dimensions,'' in
  \emph{Proceedings of the 15th ACM SIGKDD international conference on
  Knowledge discovery and data mining}.\hskip 1em plus 0.5em minus 0.4em\relax
  ACM, 2009, pp. 817--826.

\bibitem{ramos2003using}
J.~Ramos \emph{et~al.}, ``Using tf-idf to determine word relevance in document
  queries,'' in \emph{Proceedings of the first instructional conference on
  machine learning}, vol. 242.\hskip 1em plus 0.5em minus 0.4em\relax
  Piscataway, NJ, 2003, pp. 133--142.

\bibitem{schlichtkrull2018modeling}
M.~Schlichtkrull, T.~N. Kipf, P.~Bloem, R.~Van Den~Berg, I.~Titov, and
  M.~Welling, ``Modeling relational data with graph convolutional networks,''
  in \emph{Proceedings of the 15th European Semantic Web Conference}.\hskip 1em
  plus 0.5em minus 0.4em\relax Springer, 2018, pp. 593--607.

\bibitem{kingma2014adam}
D.~P. Kingma and J.~Ba, ``Adam: {A} method for stochastic optimization,'' in
  \emph{Proceedings of the 3rd International Conference on Learning
  Representations, {ICLR} 2015, San Diego, CA, USA, May 7-9, 2015, Conference
  Track Proceedings}, Y.~Bengio and Y.~LeCun, Eds., 2015.

\bibitem{xu2018powerful}
K.~Xu, W.~Hu, J.~Leskovec, and S.~Jegelka, ``How powerful are graph neural
  networks?'' \emph{Proceedings of the 7th International Conference on Learning
  Representations, {ICLR} 2019, New Orleans, LA, USA, May 6-9, 2019, Conference
  Track Proceedings}, 2019.

\end{thebibliography}

\begin{IEEEbiography}[{\includegraphics[width=25mm, clip,keepaspectratio]{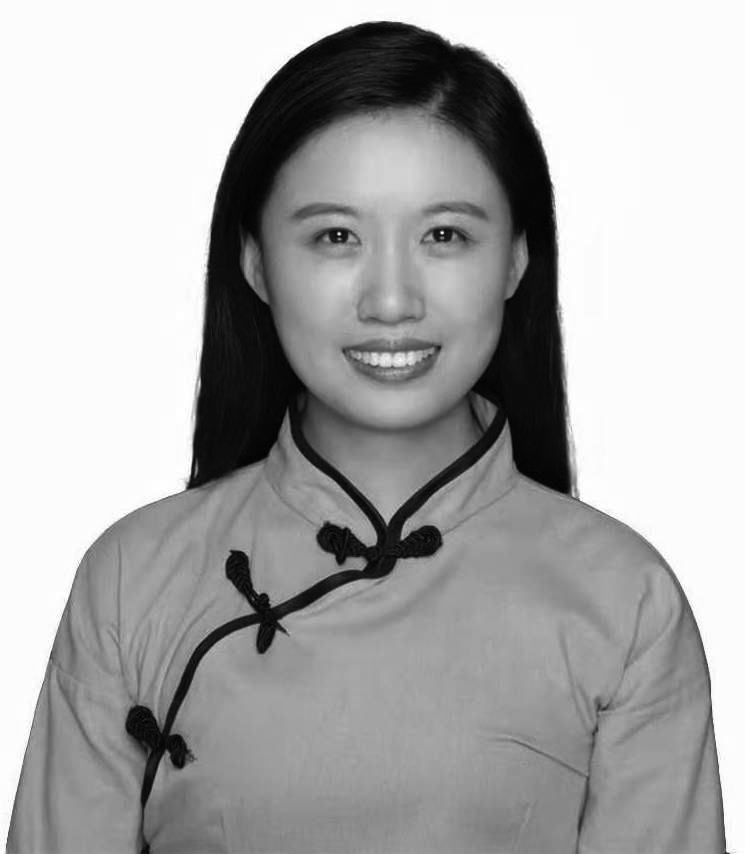}}]
{Hong Huang} is an associate professor in Huazhong University of Science and Technology, China. She received her Ph.D degree in Computer Science from University of G\"{o}ttingen, Germany in 2016, and her M.E. degree in Electronic Engineering from Tsinghua University, Beijing, China in 2012. Her research interests lie in social network analysis, data mining and knowledge graph.
\vspace{-0.2in}
\end{IEEEbiography}

\begin{IEEEbiography}[{\includegraphics[width=25mm, clip,keepaspectratio]{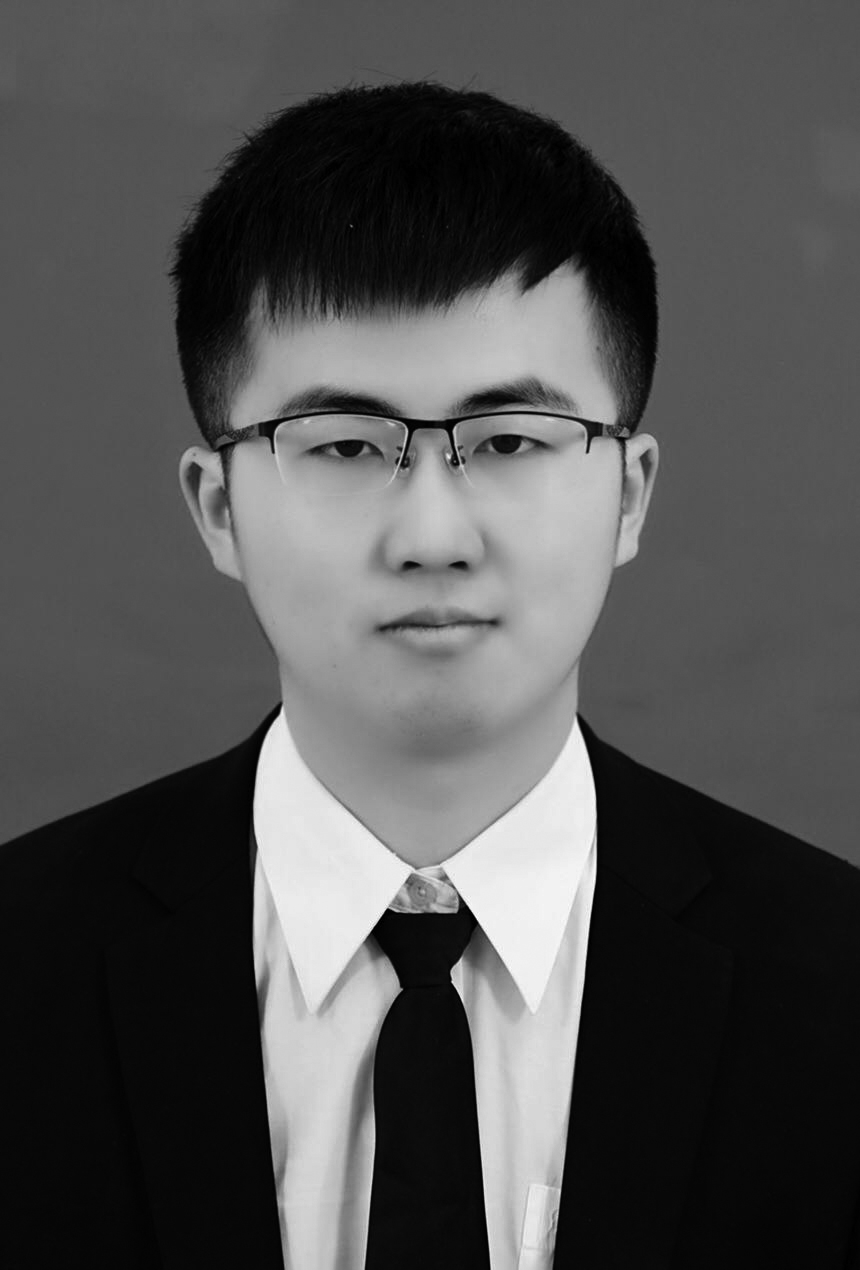}}]
{Yu Song} received his B.S. degree in Electronic Information Engineering from Huazhong University of Science and Technology, Wu Han, China, in 2018. Currently, he is pursuing his master degree in Huazhong University of Science and Technology. His research interests include data mining and social network analysis.
\end{IEEEbiography}

\begin{IEEEbiography}[{\includegraphics[width=25mm, clip,keepaspectratio]{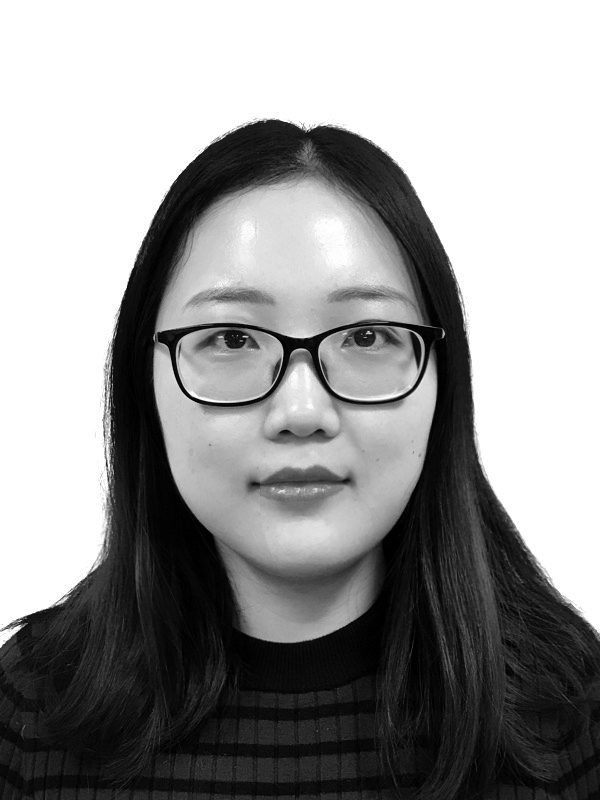}}]
{Yao Wu} received the B.S. degree in communication from Huazhong University of Science and Technology, Wuhan, China, in 2016. Currently, she is pursuing her Ph.D. degree in Huazhong University of Science and Technology. Her research interests include machine learning, data mining, social network analysis.
\end{IEEEbiography}

\begin{IEEEbiography}[{\includegraphics[width=25mm, clip,keepaspectratio]{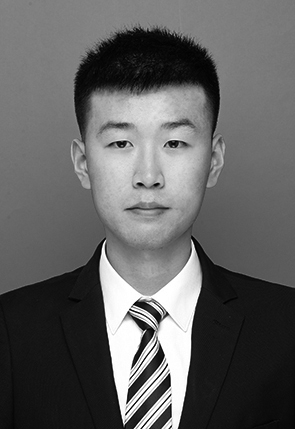}}]
{Jia Shi} received his B.S. degree from Lanzhou University, China, in 2018. Currently, he is pursuing his master degree in Huazhong University of Science and Technology. His research fields include data mining and natural language processing.
\end{IEEEbiography}

\begin{IEEEbiography}[{\includegraphics[width=25mm, clip,keepaspectratio]{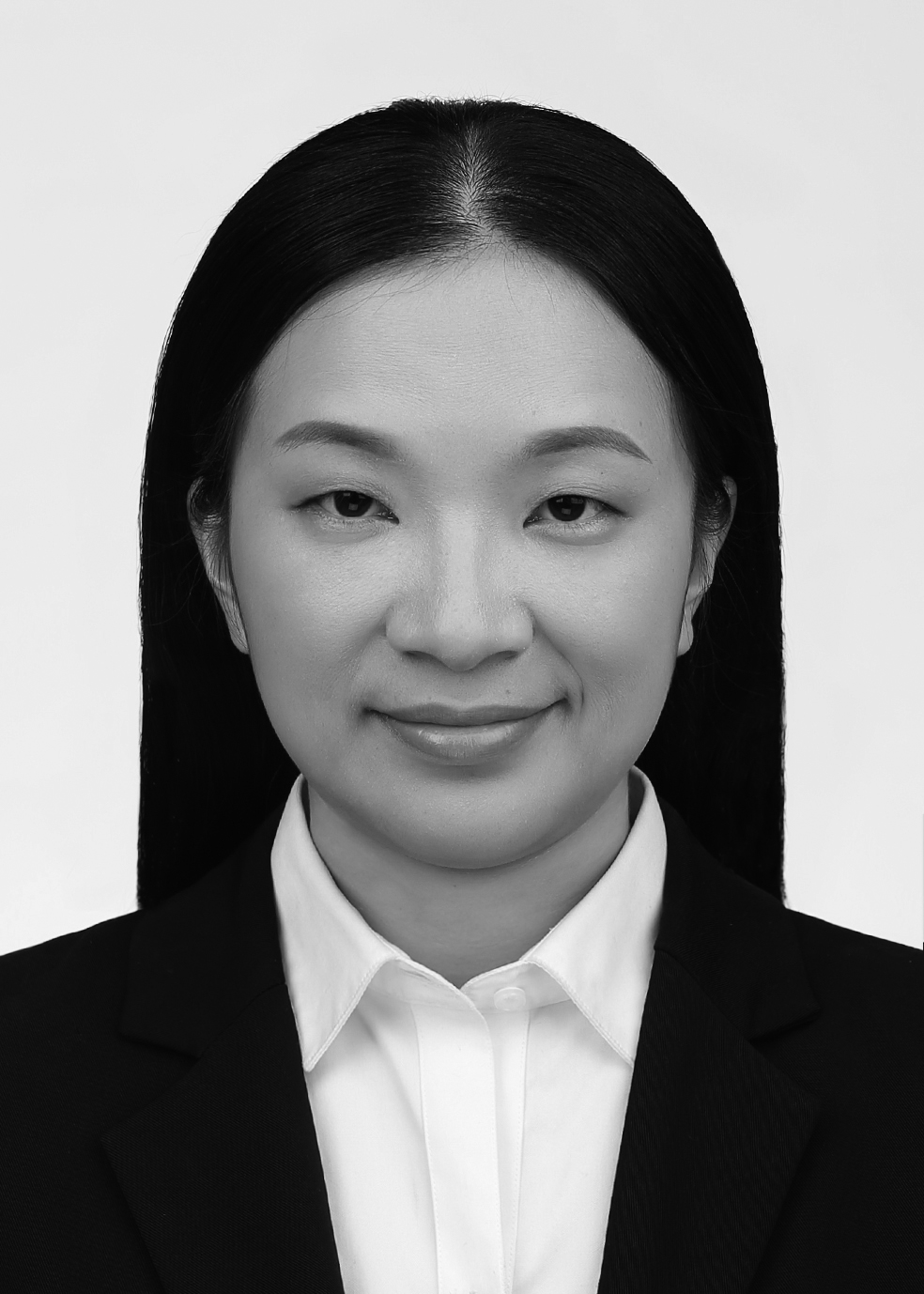}}]
{Xia Xie} is an associate professor at Huazhong University of Science and Technology (HUST) in China. She received her Ph.D. in computer architecture from HUST in 2006. Her research interests include data mining and big data.
\end{IEEEbiography}

\begin{IEEEbiography}[{\includegraphics[width=25mm, clip,keepaspectratio]{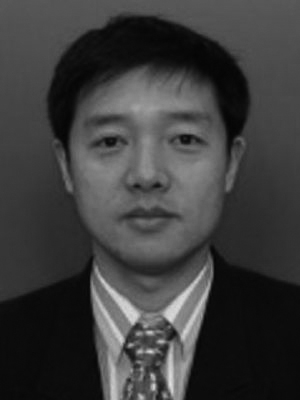}}]
{Hai Jin} is a Cheung Kung Scholars Chair Professor of computer science and engineering at Huazhong University of Science and Technology (HUST) in China. Jin received his PhD in computer engineering from HUST in 1994. In 1996, he was awarded a German Academic Exchange Service fellowship to visit the Technical University of Chemnitz in Germany. Jin worked at The University of Hong Kong between 1998 and 2000, and as a visiting scholar at the
University of Southern California between 1999 and 2000. He was awarded Excellent Youth Award from the National Science Foundation of China in 2001. Jin is the chief scientist of ChinaGrid, the largest grid computing project in China, and the chief scientists of National 973 Basic Research Program Project of Virtualization Technology of Computing System, and Cloud Security. Jin is a fellow of the IEEE, a fellow of the China Computer Federation~(CCF), and a member of the Association for Computing Machinery~(ACM). He has co-authored 22 books and published over 700 research papers. His research interests include computer architecture, virtualization technology, cluster computing and cloud computing, peer-to-peer computing, network storage, big data and network security.
\end{IEEEbiography}

\end{document}